\pdfoutput=1
\documentclass[12pt]{article}

\usepackage{cite}
\usepackage{float} 
\usepackage{graphicx}
\usepackage{amsfonts}
\usepackage{float}
\usepackage{longtable}
\usepackage{comment}

\input{epsf}

\advance \topmargin by -\headheight
\advance \topmargin by -\headsep

\evensidemargin \oddsidemargin
\begin{document}

\def\rh{{\hat \rho}}
\def\alie{{\hat{\cal G}}}
\def\hpsi{{\widehat \psi}}
\newcommand{\sect}[1]{\setcounter{equation}{0}\section{#1}}
\renewcommand{\theequation}{\thesection.\arabic{equation}}

\def\rf#1{(\ref{eq:#1})}
\def\lab#1{\label{eq:#1}}
\def\nonu{\nonumber}
\def\br{\begin{eqnarray}}
\def\er{\end{eqnarray}}
\def\be{\begin{equation}}
\def\ee{\end{equation}}
\def\eq{\!\!\!\! &=& \!\!\!\! }
\def\foot#1{\footnotemark\footnotetext{#1}}
\def\lb{\lbrack}
\def\rb{\rbrack}
\def\llangle{\left\langle}
\def\rrangle{\right\rangle}
\def\blangle{\Bigl\langle}
\def\brangle{\Bigr\rangle}
\def\llbrack{\left\lbrack}
\def\rrbrack{\right\rbrack}
\def\lcurl{\left\{}
\def\rcurl{\right\}}
\def\({\left(}
\def\){\right)}
\newcommand{\nit}{\noindent}
\newcommand{\ct}[1]{\cite{#1}}
\newcommand{\bi}[1]{\bibitem{#1}}
\def\lskip{\vskip\baselineskip\vskip-\parskip\noindent}
\relax

\def\tr{\mathop{\rm tr}}
\def\Tr{\mathop{\rm Tr}}
\def\trace{\widehat{\rm Tr}}
\def\v{\vert}
\def\bv{\bigm\vert}
\def\Bgv{\;\Bigg\vert}
\def\bgv{\bigg\vert}
\newcommand\partder[2]{{{\partial {#1}}\over{\partial {#2}}}}
\newcommand\funcder[2]{{{\delta {#1}}\over{\delta {#2}}}}
\newcommand\Bil[2]{\Bigl\langle {#1} \Bigg\vert {#2} \Bigr\rangle}  
\newcommand\bil[2]{\left\langle {#1} \bigg\vert {#2} \right\rangle} 
\newcommand\me[2]{\left\langle {#1}\bv {#2} \right\rangle} 
\newcommand\sbr[2]{\left\lbrack\,{#1}\, ,\,{#2}\,\right\rbrack}
\newcommand\pbr[2]{\{\,{#1}\, ,\,{#2}\,\}}
\newcommand\pbbr[2]{\lcurl\,{#1}\, ,\,{#2}\,\rcurl}

\def\ket#1{\mid {#1} \rangle}
\def\bra#1{\langle {#1} \mid}
\newcommand{\braket}[2]{\langle {#1} \mid {#2}\rangle}
%
\def\a{\alpha}
\def\at{{\tilde A}^R}
\def\atc{{\tilde {\cal A}}^R}
\def\atcm#1{{\tilde {\cal A}}^{(R,#1)}}
\def\b{\beta}
\def\dc{{\cal D}}
\def\d{\delta}
\def\D{\Delta}
\def\eps{\epsilon}
\def\vareps{\varepsilon}
\def\g{\gamma}
\def\G{\Gamma}
\def\grad{\nabla}
\def\h{{1\over 2}}
\def\l{\lambda}
\def\L{\Lambda}
\def\m{\mu}
\def\n{\nu}
\def\o{\over}
\def\om{\omega}
\def\O{\Omega}
\def\p{\phi}
\def\P{\Phi}
\def\pa{\partial}
\def\pr{\prime}
\def\pt{{\tilde \Phi}}
\def\qs{Q_{\bf s}}
\def\ra{\rightarrow}
\def\s{\sigma}
\def\S{\Sigma}
\def\t{\tau}
\def\th{\theta}
\def\Th{\Theta}
\def\tpp{\Theta_{+}}
\def\tmm{\Theta_{-}}
\def\tpg{\Theta_{+}^{>}}
\def\tms{\Theta_{-}^{<}}
\def\tp0{\Theta_{+}^{(0)}}
\def\tm0{\Theta_{-}^{(0)}}
\def\ti{\tilde}
\def\wti{\widetilde}
\def\jc{J^C}
\def\bj{{\bar J}}
\def\sj{{\jmath}}
\def\bsj{{\bar \jmath}}
\def\bp{{\bar \p}}
\def\vp{\varphi}
\def\ve{\varepsilon}
\def\vt{{\tilde \varphi}}
\def\faa{Fa\'a di Bruno~}
\def\ca{{\cal A}}
\def\cb{{\cal B}}
\def\ce{{\cal E}}
\def\cg{{\cal G}}
\def\cgh{{\hat {\cal G}}}
\def\ch{{\cal H}}
\def\chh{{\hat {\cal H}}}
\def\cl{{\cal L}}
\def\cm{{\cal M}}
\def\cn{{\cal N}}
\def\u2{\mid u\mid^2}
\def\ub{{\bar u}}
\def\z2{\mid z\mid^2}
\def\zb{{\bar z}}
\def\w2{\mid w\mid^2}
\def\wb{{\bar w}}
\newcommand\sumi[1]{\sum_{#1}^{\infty}}   
\newcommand\fourmat[4]{\left(\begin{array}{cc}  
{#1} & {#2} \\ {#3} & {#4} \end{array} \right)}

%
\def\lie{{\cal G}}
\def\kmlie{{\hat{\cal G}}}
\def\dlie{{\cal G}^{\ast}}
\def\elie{{\widetilde \lie}}
\def\edlie{{\elie}^{\ast}}
\def\hlie{{\cal H}}
\def\flie{{\cal F}}
\def\wlie{{\widetilde \lie}}
\def\f#1#2#3 {f^{#1#2}_{#3}}
\def\winf{{\sf w_\infty}}
\def\win1{{\sf w_{1+\infty}}}
\def\hwinf{{\sf {\hat w}_{\infty}}}
\def\Winf{{\sf W_\infty}}
\def\Win1{{\sf W_{1+\infty}}}
\def\hWinf{{\sf {\hat W}_{\infty}}}
\def\Rm#1#2{r(\vec{#1},\vec{#2})}          
\def\OR#1{{\cal O}(R_{#1})}           
\def\ORti{{\cal O}({\widetilde R})}           
\def\AdR#1{Ad_{R_{#1}}}              
\def\dAdR#1{Ad_{R_{#1}^{\ast}}}      
\def\adR#1{ad_{R_{#1}^{\ast}}}       
\def\KP{${\rm \, KP\,}$}                 
\def\KPl{${\rm \,KP}_{\ell}\,$}         
\def\KPo{${\rm \,KP}_{\ell = 0}\,$}         
\def\mKPa{${\rm \,KP}_{\ell = 1}\,$}    
\def\mKPb{${\rm \,KP}_{\ell = 2}\,$}    
%
\def\rlx{\relax\leavevmode}
\def\inbar{\vrule height1.5ex width.4pt depth0pt}
\def\IZ{\rlx\hbox{\sf Z\kern-.4em Z}}
\def\IR{\rlx\hbox{\rm I\kern-.18em R}}
\def\IC{\rlx\hbox{\,$\inbar\kern-.3em{\rm C}$}}
\def\IN{\rlx\hbox{\rm I\kern-.18em N}}
\def\IO{\rlx\hbox{\,$\inbar\kern-.3em{\rm O}$}}
\def\IP{\rlx\hbox{\rm I\kern-.18em P}}
\def\IQ{\rlx\hbox{\,$\inbar\kern-.3em{\rm Q}$}}
\def\IF{\rlx\hbox{\rm I\kern-.18em F}}
\def\IG{\rlx\hbox{\,$\inbar\kern-.3em{\rm G}$}}
\def\IH{\rlx\hbox{\rm I\kern-.18em H}}
\def\II{\rlx\hbox{\rm I\kern-.18em I}}
\def\IK{\rlx\hbox{\rm I\kern-.18em K}}
\def\IL{\rlx\hbox{\rm I\kern-.18em L}}
\def\one{\hbox{{1}\kern-.25em\hbox{l}}}
\def\0#1{\relax\ifmmode\mathaccent"7017{#1}%
B        \else\accent23#1\relax\fi}
\def\omz{\0 \omega}
%
\def\ltimes{\mathrel{\vrule height1ex}\joinrel\mathrel\times}
\def\rtimes{\mathrel\times\joinrel\mathrel{\vrule height1ex}}
%
\def\mark{\noindent{\bf Remark.}\quad}
\def\prop{\noindent{\bf Proposition.}\quad}
\def\theor{\noindent{\bf Theorem.}\quad}
\def\name{\noindent{\bf Definition.}\quad}
\def\exam{\noindent{\bf Example.}\quad}
\def\proof{\noindent{\bf Proof.}\quad}
\newcommand{\map}{\mathcal{P}}

\begin{titlepage}
\vspace*{-1cm}

\vskip 3cm

\vspace{.2in}
\begin{center}
{\large\bf  Harmonic, Holomorphic and Rational Maps from  Self-Duality }
\end{center}

\vspace{.5cm}

\begin{center}

L. A. Ferreira$^{\dagger ,}$\footnote{laf@ifsc.usp.br} and L. R. Livramento$^{\dagger,}$\footnote{livramento@usp.br}

\vspace{.3 in}
\small

\par \vskip .2in \noindent
$^{\dagger}$Instituto de F\'\i sica de S\~ao Carlos; IFSC/USP;\\
Universidade de S\~ao Paulo, USP  \\ 
Caixa Postal 369, CEP 13560-970, S\~ao Carlos-SP, Brazil\\

\vskip 2cm

\begin{abstract}

We propose a generalization of the so-called rational map ansatz on the Euclidean space  $\IR^3$, for any compact simple Lie group $G$ such that $G/{\widehat K}\otimes U(1)$ is an Hermitian symmetric space, for some subgroup ${\widehat K}$ of $G$. It generalizes the rational maps on the two-sphere $SU(2)/U(1)$, and also on $CP^N=SU(N+1)/SU(N)\otimes U(1)$, and opens up the way for applications of such ans\"atze on non-linear sigma models, Skyrme theory and magnetic monopoles in Yang-Mills-Higgs theories. Our construction is based on a well known mathematical result stating that  stable harmonic maps $X$ from the two-sphere $S^2$ to compact Hermitian symmetric spaces $G/{\widehat K}\otimes U(1)$ are holomorphic or anti-holomorphic. We derive such a mathematical result using ideas involving the concept of self-duality, in a way that makes it more accessible to theoretical physicists. Using a to\-po\-lo\-gi\-cal (homotopic) charge that admits an integral representation, we construct first order partial differential self-duality equations such that their solutions also solve the (second order) Euler-Lagrange associated to the  harmonic map energy  $E=\int_{S^2} \mid dX\mid^2 d\mu$. We show that such solutions saturate a lower bound on the energy $E$, and that the self-duality equations constitute the Cauchy-Riemann equations for the maps $X$. Therefore, they constitute harmonic  and (anti)holomorphic maps, and lead to the generalization of the rational map ans\"atze in $\IR^3$.  We apply our results to construct approximate Skyrme solutions for the $SU(N)$ Skyrme model.

\end{abstract}

\normalsize
\end{center}
\end{titlepage}

\section{Introduction}
\label{sec:introduction}
\setcounter{equation}{0}

The construction of exact or approximate solutions of differential equations are greatly simplified when the order and/or the dimensions of those equations can be reduced. In general, such a reduction is achieved by the use of ans\"atze based on the symmetries of the equations, and the Lie method is the prototype of it. However, there are cases where  the topological and algebraic  structures, underlying the problem under investigation, can be of great help. An example of that is the so-called rational map ans\"atze. It has been widely used in many applications in topological solitons \cite{mantonbook}, like the lumps in $(2+1)$-dimensions \cite{belavincp1}, in the Naham data for self-dual magnetic monopoles in Yang-Mills-Higgs theories in $(3+1)$-dimensions \cite{donaldson,jarvis}, and to provide  good approximations for solutions of the Skyrme model \cite{rational1,rational2,mantonbook}. 

In the case of the Skyrmions  most applications are in the context of $SU(2)$ Skyrme model, where the rational maps are holomorphic functions from two-sphere $S^2$ to the complex projective space $CP^1=SU(2)/U(1)$. The rational map ansatz provides an elegant geometric construction that effectively captures most of the key features of Skyrmions, including an accurate approximation of their static energy. There exists a generalization for the $SU(N+1)$ Skyrme model with rational maps from $S^2$ to $CP^N=SU(N+1)/SU(N)\otimes U(1)$ \cite{ioannidou1,ioannidou2,wojtekbook}. 

In this paper we generalize the rational map ans\"atze on the Euclidean space  $\IR^3$, for any compact simple Lie group $G$ such that $G/{\widehat K}\otimes U(1)$ is a compact Hermitian symmetric space, for some subgroup ${\widehat K}$ of $G$.  For a given element $U\in G$, the ansatz has the form
\be
U= e^{i\,f\(r\)\,g\,\Lambda\,g^{-1}}
\lab{hologintro}
\ee
where $f\(r\)$ is a radial profile function, $\Lambda$ is the generator of the $U(1)$ subgroup of $G$ appearing in Hermitian symmetric space $G/{\widehat K}\otimes U(1)$, and $g$ is a matrix (principal variable) that parameterizes the coset $G/{\widehat K}\otimes U(1)$. We use the spherical coordinates in $\IR^3$, $\(r\,,\,\, z\,,\, {\bar z}\)$, where $r$ is the radial coordinate, and the spheres of a given radius $r$ are stereographically projected on the complex plane $z$.  The matrices $g$ and $g\,\Lambda\,g^{-1}$ are given by 
\be
g\equiv e^{i\,S}\,e^{\varphi\,\sbr{S}{S^{\dagger}}}\,e^{i\,S^{\dagger}} \,; \qquad\qquad g\,\Lambda\,g^{-1}=\Lambda -\frac{1}{\(1+\omega\)}\left(\sbr{S}{S^{\dagger}}+i\(S-S^{\dagger}\)\right)
\lab{gdefintro}
\ee
with $S$ being a matrix in some special representation of $G$ such that
\be
S^2=0\;;\qquad\qquad \qquad \qquad\(S\,S^{\dagger}\)\,S=\omega\,S
\lab{omegadefintro}
\ee
with $\omega$ a real and non-negative eigenvalue, and $\varphi=\frac{\ln\sqrt{1+\omega}}{\omega}$. The matrix $S$ is either holomorphic, $\partial_{{\bar z}} S=0$, or anti-holomorphic, $\partial_{z} S=0$. Therefore, the matrix $g$ defines  holomorphic, or anti-holomorphic, maps from the two-spheres $S^2$ in $\IR^3$, parameterized by  $z$ and ${\bar z}$, to the Hermitian symmetric space $G/{\widehat K}\otimes U(1)$.  Such maps are also harmonic since $g$ saturates the lower bound on the standard energy of the non-linear sigma model on $G/{\widehat K}\otimes U(1)$. 

We have that the second homotopy group of any Hermitian symmetric space is the integers, i.e. $\pi_2\(G/{\widehat K}\otimes U(1)\)=\IZ$. In addition, the third homotopy group of any simple compact Lie group is also the integers, i.e. $\pi_3\(G\)=\IZ$. Thefore, for those configurations where the group element $U$ goes to a constant at spatial infinity in $\IR^3$, the ansatz \rf{hologintro} gives a map which is an element of $\pi_3\(G\)=\IZ$, with degree $Q$ given by 
\be
Q=\frac{N}{2\,\pi}\,\left[f\(r\)-\sin f\(r\)\right]_{r=0}^{r=\infty} 
\lab{charge3dintro}
\ee
with $N$ being the degree of the (anti)holomorphic and harmonic map, $S^2\rightarrow G/{\widehat K}\otimes U(1)$, given by $g$. 

Our construction is based on  well known mathematical results due to Lichnerowicz \cite{lic69},  Burstall, Rawnsley and Salamon \cite{brs,burstall1987}, and Eells and Lemaire \cite{eells}. In a succinct form, those results state that  any stable harmonic map of a compact Riemann surface into a  compact simple Hermitian symmetric space  is holomorphic or anti-holomorphic \cite{eells}. We derive those well known mathematical results in a form which is accessible to theoretical physicists. 

The starting point are the self-duality equations for the $(2+1)$-dimensional non-linear sigma model on the Hermitian symmetric space $G/{\widehat K}\otimes U(1)$  \cite{lic69}. We use a generalization of the concept of self-duality put forward in \cite{genbps} which allows the use of complex fields. Starting from an  integral representation of the topological charge, which is the degree of the maps $S^2\rightarrow G/{\widehat K}\otimes U(1)$, we derive the self-duality first order  differential equations. The static solutions of such self-duality equations also solve the Euler-Lagrange second order differential equations of the non-linear model on $G/{\widehat K}\otimes U(1)$. In addition, such static self-dual solutions saturate a lower bound, given by the topological charge, on the static energy of that model. Therefore, the self-dual solutions provide harmonic maps from $S^2$, the compactified two dimensional plane, to the Hermitian symmetric space $G/{\widehat K}\otimes U(1)$. It turns out that the self-duality equations become the Cauchy-Riemann equations for the matrix $S$ defined in \rf{omegadefintro}, and therefore the maps $S^2\rightarrow G/{\widehat K}\otimes U(1)$ are either holomorphic or anti-holomorphic. The construction of the rational map ansatz \rf{hologintro} then follows in a quite direct way from harmonic and (anti)holomorphic maps described above.

Our generalized rational map ansatz is a powerful tool for constructing topological solitons in nonlinear sigma models. Although it may not lead to solutions of all the Euler-Lagrange equations in such theories, energy minimization within this ansatz may provide good approximations for the global energy minimizers, in addition to giving an upper energy bound for those solutions. In fact, a well-known feature of the standard $SU(2)$ Skyrme model \cite{skyrme1, skyrme2, mantonbook} is that the usual rational map ansatz leads to the global energy minimizer for $Q=1$ and provides a very good approximation of the energy minimizers for $Q \geq 2$. On the other hand, the ansatz can serve as a good starting point for applying fully 3D minimization methods capable of obtaining global energy minimizers.

The paper is organized as follows: in Section \ref{sec:self-duality} we give a generalization of the concept of self-duality, proposed \cite{genbps}, to the case of complex fields. In Section \ref{sec:symmetric} we give the algebraic construction of the compact simple Hermitian symmetric spaces and the complete list of them. The construction of the harmonic and (anti)holomorphic ans\"atze is given in Section \ref{sec:harmoholo}. The three dimensional rational map \rf{hologintro} is given in section \ref{sec:rationalmap}, and an explicit construction of the matrices $g$ and $S$, introduced in \rf{gdefintro} and \rf{omegadefintro} respectively, is provided. Then in Section \ref{sec:examples} we construct explicit examples of our ansatz  \rf{hologintro} for the Hermitian symmetric spaces $CP^1=SU(2)/U(1)$, $CP^N=SU(N+1)/SU(N)\otimes U(1)$, $SU(p+q)/SU(p)\otimes SU(q)\otimes U(1)$, and $Sp(N)/SU(N)\otimes U(1)$. In Section \ref{sec:applications} we apply our ansatz to construct approximate Skyrme solutions for the $SU(N)$ Skyrme model, and in in Section \ref{sec:conclusion}  we present our conclusion.

\section{The concept of self-duality}
\label{sec:self-duality}
\setcounter{equation}{0}

The concept of self-duality has been used for a long time in several contexts \cite{bogo,prasad,belavincp1,belavininstanton}, and we give here the main idea behind the concept of generalized self-duality proposed in \cite{genbps}, extending it to the case of complex fields.  Consider a field theory that possesses a topological charge with an integral representation of the form
\be
Q=\frac{1}{2}\,\int d^dx\,\left[ {\cal A}_{\alpha}\,{\widetilde{\cal A}}_{\alpha}^*+{\cal A}_{\alpha}^*\,{\widetilde{\cal A}}_{\alpha}\right]
\lab{topcharge}
\ee
where  ${\cal A}_{\alpha}$ and ${\widetilde{\cal A}}_{\alpha}$ are functionals of the fields of the theory and their first derivatives only, and where $^*$ means complex conjugation, and not transpose complex conjugate.  The index $\alpha$ stands for any type of indices, like vector, spinor, internal, etc, or sets of them, and summation over $\alpha$ is implied whenever we have repeated indices. The fact that $Q$ is topological means that it is invariant under any smooth (homotopic) variation of the fields. Let us denote the fields by $\chi_{\kappa}$, and they can be scalar, vector, spinor fields, and the index $\kappa$ stands for the space-time and internal indices. We take $\chi_{\kappa}$ to be real, and so, if there are complex fields, $\chi_{\kappa}$ stands for their real and imaginary parts. The invariance of $Q$ under smooth variations of the fields lead to the identities
\br
\delta\,Q=0\quad\rightarrow \quad &&
\frac{\delta\, {\cal A}_{\alpha}}{\delta\,\chi_{\kappa}}\, {\widetilde{\cal A}}_{\alpha}^*-\partial_{\mu}\(\frac{\delta\, {\cal A}_{\alpha}}{\delta\,\partial_{\mu}\chi_{\kappa}}\, {\widetilde{\cal A}}_{\alpha}^*\)
+
{\cal A}_{\alpha}\,\frac{\delta\, {\widetilde{\cal A}}_{\alpha}^*}{\delta\,\chi_{\kappa}} -\partial_{\mu}\({\cal A}_{\alpha}\,\frac{\delta\, {\widetilde{\cal A}}_{\alpha}^*}{\delta\,\partial_{\mu}\chi_{\kappa}}\)
\lab{topidentity}\\
&&
\frac{\delta\, {\cal A}_{\alpha}^*}{\delta\,\chi_{\kappa}}\, {\widetilde{\cal A}}_{\alpha}-\partial_{\mu}\(\frac{\delta\, {\cal A}_{\alpha}^*}{\delta\,\partial_{\mu}\chi_{\kappa}}\, {\widetilde{\cal A}}_{\alpha}\)
+
{\cal A}_{\alpha}^*\,\frac{\delta\, {\widetilde{\cal A}}_{\alpha}}{\delta\,\chi_{\kappa}} -\partial_{\mu}\({\cal A}_{\alpha}^*\,\frac{\delta\, {\widetilde{\cal A}}_{\alpha}}{\delta\,\partial_{\mu}\chi_{\kappa}}\)
=0
\nonumber
\er
By imposing the first order differential equations, or self-duality equations, on the fields as
\be
{\cal A}_{\alpha}=\pm {\widetilde{\cal A}}_{\alpha}
\lab{sdeqs}
\ee
it follows that, together with the identities \rf{topidentity}, they imply the equations 
\br
&&\frac{\delta\, {\cal A}_{\alpha}}{\delta\,\chi_{\kappa}}\, {{\cal A}}_{\alpha}^*-\partial_{\mu}\(\frac{\delta\, {\cal A}_{\alpha}}{\delta\,\partial_{\mu}\chi_{\kappa}}\, {{\cal A}}_{\alpha}^*\)
+
 {{\cal A}}_{\alpha}\,\frac{\delta\, {\cal A}_{\alpha}^*}{\delta\,\chi_{\kappa}}-\partial_{\mu}\( {{\cal A}}_{\alpha}\,\frac{\delta\, {\cal A}_{\alpha}^*}{\delta\,\partial_{\mu}\chi_{\kappa}}\)
 \lab{eleqs}\\
&& +
\frac{\delta\, {\widetilde{\cal A}}_{\alpha}^*}{\delta\,\chi_{\kappa}}\,{\widetilde{\cal A}}_{\alpha} -\partial_{\mu}\(\frac{\delta\, {\widetilde{\cal A}}_{\alpha}^*}{\delta\,\partial_{\mu}\chi_{\kappa}}\,{\widetilde{\cal A}}_{\alpha}\)
+
{\widetilde{\cal A}}_{\alpha}^*\,\frac{\delta\, {\widetilde{\cal A}}_{\alpha}}{\delta\,\chi_{\kappa}} -\partial_{\mu}\({\widetilde{\cal A}}_{\alpha}^*\,\frac{\delta\, {\widetilde{\cal A}}_{\alpha}}{\delta\,\partial_{\mu}\chi_{\kappa}}\)=0
\nonumber
\er
Note that \rf{eleqs} are the Euler-Lagrange equations associated to the functional 
\be
E=\frac{1}{2}\,\int d^dx\,\left[{\cal A}_{\alpha}\,{\cal A}_{\alpha}^*+{\widetilde{\cal A}}_{\alpha}\,{\widetilde{\cal A}}_{\alpha}^*\right]
\lab{energy}
\ee
So, first order differential equations together with second order topological identities lead to second order Euler-Lagrange equations. The fact that we have to perform one integration less to solve the Euler-Lagrange equations associated to \rf{energy} is not related to any dynamical conservation law but to the homotopic  invariance of the topological charge \rf{topcharge}. 

Note that, if $E$ is positive definite then the self-dual solutions saturate a lower bound on $E$ as follows. From \rf{sdeqs} we have that ${\cal A}_{\alpha}^2={\widetilde{\cal A}}_{\alpha}^2=\pm {\cal A}_{\alpha}\,{\widetilde{\cal A}}_{\alpha}$. Note that \rf{sdeqs} also implies that ${\cal A}_{\alpha}\,{\widetilde{\cal A}}_{\alpha}^*={\cal A}_{\alpha}^*\,{\widetilde{\cal A}}_{\alpha}$. Therefore, if ${\cal A}_{\alpha}\,{\cal A}_{\alpha}^*\geq 0$, and consequently ${\widetilde{\cal A}}_{\alpha}\,{\widetilde{\cal A}}_{\alpha}^*\geq 0$, we have that 
\br
{\cal A}_{\alpha}= {\widetilde{\cal A}}_{\alpha}\quad &\rightarrow& \quad Q=\int d^dx\, {\cal A}_{\alpha}\,{\cal A}_{\alpha}^*\geq 0
\nonumber\\
{\cal A}_{\alpha}= -{\widetilde{\cal A}}_{\alpha}\quad &\rightarrow& \quad Q=-\int d^dx\, {\cal A}_{\alpha}\,{\cal A}_{\alpha}^*\leq 0
\er
Therefore we have that 
\be
E=\frac{1}{2}\,\int d^dx\,\left[{\cal A}_{\alpha} \mp {\widetilde{\cal A}}_{\alpha}\right] \left[{\cal A}_{\alpha}^* \mp {\widetilde{\cal A}}_{\alpha}^*\right]\pm \frac{1}{2}\,\int d^dx\,\left[{\cal A}_{\alpha} \, {\widetilde{\cal A}}_{\alpha}^*+{\cal A}_{\alpha}^* \, {\widetilde{\cal A}}_{\alpha}\right]\geq \mid Q\mid
\lab{bound}
\ee
and the equality holds true for self-dual solutions, where we have
\be
E=\int d^dx\,{\cal A}_{\alpha}\,{\cal A}_{\alpha}^*=\int d^dx\,{\widetilde{\cal A}}_{\alpha}\,{\widetilde{\cal A}}_{\alpha}^*= \mid Q\mid
\ee

Summarizing, self-duality equations are first order (partial) differential equations which solutions solves second order Euler-Lagrange equations associated to an energy functional. When that energy is positive definite, the self-dual solutions saturates a lower bound on such an energy determined by the value of the topological charge. The fact that we can solve the Euler-Lagrange equations, performing one integration less, is due to  a topological charge that possesses an integral representation, and so, a topological charge density. The homotopic invariance of such a charge leads to local identities in the form of second order (partial) differential equations, that together with the first order self-duality equations, lead to solution of the Euler-Lagrange equations. 

We shall now show that self-duality equations are responsible for the existence of harmonic and holomorphic maps from the two-sphere $S^2$ to Hermitian symmetric spaces. 

\section{Hermitian Symmetric Spaces}
\label{sec:symmetric}
\setcounter{equation}{0}

\begin{figure}[htp]
\begin{center}
		\includegraphics[scale=0.5]{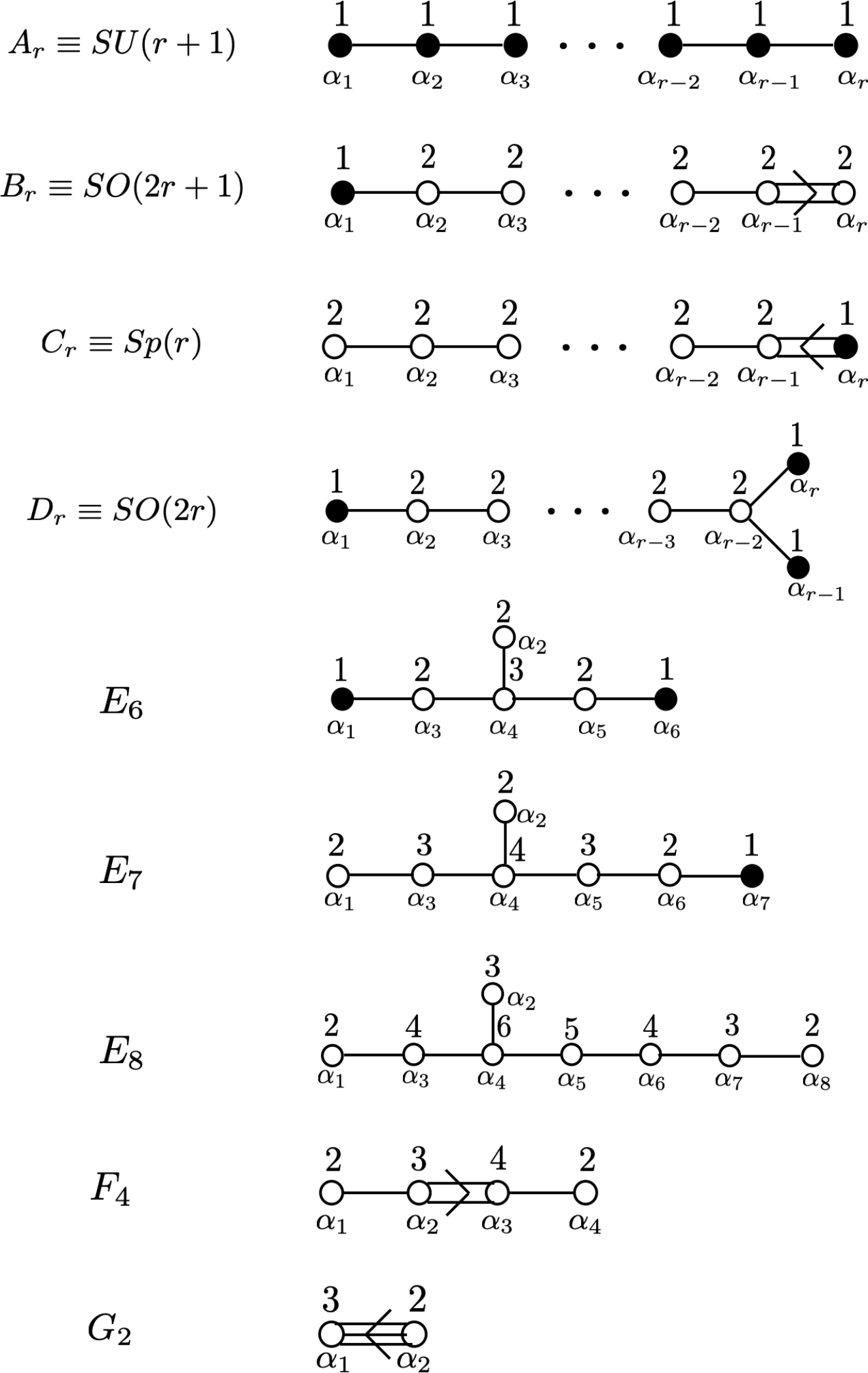}
			\caption{The Dynkin diagrams of the simple Lie algebras. The $\alpha_a$'s below the spots label the simple roots, and the numbers above correspond to the integers $m_a$ in the expansion of the highest root $\psi=\sum_{a=1}^r m_a\,\alpha_a$. The black spots correspond to $m_a=1$. Note that for $E_8$, $F_4$ and $G_2$, none of the $m_a$'s equal unity, and so such groups do not lead to Hermitian symmetric spaces. }
		\label{fig:dynkin}
\end{center} 
\end{figure}

We shall consider irreducible compact Hermitian symmetric spaces \cite{helgason}. We give here an algebraic construction of such spaces. Consider a compact simple Lie group $G$, and let $\psi$ denote its highest positive root.  The Hermitian symmetric spaces correspond to those cases where the  expansion of $\psi$ in terms of  the simple roots $\alpha_a$, $a=1,2,3\ldots {\rm rank}\,G$, presents at least one coefficient equals to unity, i.e. 
\be
\psi = \alpha_{*}+\sum_{a=1\,,\, a\neq *}^{{\rm rank}\,G}n_a\,\alpha_a\;\qquad\qquad \qquad n_a\in \IZ_{+}
\lab{alphastardef}
\ee
where we have denoted $\alpha_{*}$ that simple root that appears only once in the expansion, and all $n_a$'s are positive integers. We give in Figure \ref{fig:dynkin} the expansion $\psi=\sum_{a=1}^r m_a\,\alpha_a$, for the simple Lie algebras. Let us denote $\lambda_{*}$ the fundamental weight of $G$ which is not orthogonal to $\alpha_{*}$, i.e.
\be
\frac{2\,\lambda_{*}\cdot \alpha_{*}}{\alpha_{*}^2}=1\;;\qquad\qquad \qquad \frac{2\,\lambda_{*}\cdot \alpha_{a}}{\alpha_{a}^2}=0\;;\quad {\rm for}\quad a\neq * \lab{orthogonal}
\ee

We build an involutive inner automorphism for the Lie algebra ${\cal G}$ of $G$ as
\be
\sigma\(T\)\equiv e^{i\,\pi\,\Lambda}\, T\, e^{-i\,\pi\,\Lambda}\;;\qquad \qquad \Lambda\equiv\frac{2\,\lambda_{*}\cdot H}{\alpha_{*}^2}\;;\qquad \qquad \mbox{\rm for any}\;T\in {\cal G}
\lab{sigmadef}
\ee
where $H_{i}$, $i=1,2,3\ldots {\rm rank}\;G$, are the generators of the Cartan subalgebra of ${\cal G}$, in the Cartan-Weyl basis, and we shall denote $E_{\alpha}$ , the step operator associated to the root $\alpha$ of ${\cal G}$. The Killing form of ${\cal G}$ in such a basis is given by 
\be
{\rm Tr}\(H_i\,H_j\)=\delta_{ij}\,, \qquad\quad{\rm Tr}\(H_i\,E_{\alpha}\)=0\,,\qquad\quad{\rm Tr}\(E_{\alpha}\,E_{\beta}\)=\frac{2}{\alpha^2}\,\delta_{\alpha+\beta\,,\,0}
\lab{killingform}
\ee
The eigenvalues of $\Lambda$ in ${\cal G}$, under the adjoint action, are integers, and so $\sigma^2=1$. Therefore, $\sigma$ splits ${\cal G}$ into even and odd subspaces 
\be
{\cal G}= {\cal P}+ {\cal K}\qquad\qquad {\rm with}\qquad \sigma\(P\) = -P\qquad \sigma\(K\)=K\qquad P\in {\cal P}\;;\quad K\in {\cal K}
\lab{splittingcalg}
\ee
satisfying
\be
\sbr{{\cal K}}{{\cal K}}\subset {\cal K}\qquad\qquad\sbr{{\cal K}}{{\cal P}}\subset {\cal P}
\qquad\qquad\sbr{{\cal P}}{{\cal P}}\subset {\cal K} \lab{comut0}
\ee 
As the highest root $\psi$ contains $\alpha_{*}$ only once in its expansion, then it follows that any positive root of ${\cal G}$, either does not contain $\alpha_{*}$ in its expansion, or contains it only once. Therefore, we have
\be
\sigma\(H_i\)=H_i\;;\qquad\qquad \sigma\(E_{\pm \gamma}\)=E_{\pm \gamma}\;;\qquad\qquad \sigma\(E_{\pm \alpha_{\kappa}}\)=-E_{\pm \alpha_{\kappa}}
\lab{evenoddgenerators}
\ee
where $\gamma$ does not contain $\alpha_{*}$ in its expansion in terms of simple roots, and $\alpha_{\kappa}=\alpha_{*}+\beta_{\kappa}$,  with $\beta_{\kappa}$ being an integer linear combination of simple roots other than  $\alpha_{*}$. Obviously, $\Lambda$ belongs to ${\cal K}$, and it generates an $U(1)_{\Lambda}$ invariant subalgebra of it. Therefore, ${\cal K}={\widehat {\cal K}}\oplus {\Lambda}$, and we obtain the irreducible compact Hermitian symmetric space $G/{\widehat K}\otimes U(1)_{\Lambda}$. The subgroup ${\widehat K}$ is generated by $H_a\equiv \frac{2\,\alpha_a\cdot H}{\alpha_a^2}$, with $\alpha_a\neq \alpha_{*}$,  $\(E_{\gamma}+E_{-\gamma}\)$, and $i\,\(E_{\gamma}-E_{-\gamma}\)$, with $\gamma$ being positive roots of $G$ not containing $\alpha_{*}$ in their expansions in terms of simple roots. The odd generators are $E_{\pm \alpha_{\kappa}}$, as defined in \rf{evenoddgenerators},  and ${\rm dim}\;{\cal P}$ is clearly even, and so $\kappa=1,2\ldots \frac{{\rm dim}\;{\cal P}}{2}$.

The hermitian character of such symmetric spaces is that ${\cal P}$ is even dimensional and it is split by $\Lambda$ into two parts according to its eigenvalues
\be
{\cal P}={\cal P}_{+}+{\cal P}_{-}\qquad\qquad \sbr{\Lambda}{P_{\pm}}=\pm P_{\pm}\qquad\qquad P_{\pm}\in {\cal P}_{\pm}
\lab{hermitianniceprop}
\ee
with ${\cal P}_{+}$ being generated by $E_{\alpha_{\kappa}}$, and ${\cal P}_{-}$  by $E_{-\alpha_{\kappa}}$. Since $G$ is compact, any finite representation is equivalent to a unitary one. So, we can always have the hermiticity condition  $E_{\alpha_{\kappa}}^{\dagger}=E_{-\alpha_{\kappa}}$. Consequently, one can consider ${\cal P}_{-}$ as the hermitian conjugate of ${\cal P}_{+}$. Therefore, $\Lambda$ not only provides the automorphism $\sigma$, but it also provides a gradation of the Lie algebra ${\cal G}$ into subspaces of grades $0$ and $\pm 1$. Since there are no subspaces of grades $\pm 2$, it turns out that ${\cal P}_{\pm}$ are abelian. So we have
\be
{\cal G}={\cal G}_{-1}+{\cal G}_0+{\cal G}_{+1};\qquad \qquad \sbr{\Lambda}{{\cal G}_{n}}=n\,{\cal G}_{n}
\lab{grading}
\ee
with ${\cal G}_0\equiv {\cal K}$, and ${\cal G}_{\pm 1}\equiv {\cal P}_{\pm}$. In addition, we have 
\be
 \sbr{{\cal K}}{{\cal K}}\subset {\cal K}\qquad  \sbr{{\cal K}}{{\cal P}_{\pm}}\subset {\cal P}_{\pm}\qquad \sbr{{\cal P}_{+}}{{\cal P}_{-}}\subset {\cal K}\qquad 
 \sbr{{\cal P}_{+}}{{\cal P}_{+}}=\sbr{{\cal P}_{-}}{{\cal P}_{-}}=0
 \lab{finerstructure}
 \ee
 and from \rf{killingform}
 \be
 {\rm Tr}\({\cal K}\, {\cal P}_{\pm}\)={\rm Tr}\({\cal P}_{+}\, {\cal P}_{+}\)={\rm Tr}\({\cal P}_{-}\, {\cal P}_{-}\)=0
 \lab{granding2}
 \ee
 Following the expansion of the highest root $\psi$ given in Figure \ref{fig:dynkin} we get the following compact irreducible Hermitian symmetric spaces shown in Table \ref{listhermitian}.
 
 \begin{table}
 \begin{tabular}{ll}
 { Type I:}  & $SU(p+q)/SU(p)\otimes SU\(q\)\otimes U(1)$, by taking $\alpha_{*}$ to be any of the simple \\
 & roots for the case $G=SU(r+1)$. \\
  { Type II:}  & $SO\(N\)/SO\(N-2\)\otimes U\(1\)$, by taking $\alpha_{*}=\alpha_1$ for the cases \\
  & $G=SO\(2r+1\)$ and $G=SO\(2r\)$.\\
  { Type III:}  & $SO\(2r\)/SU\(r\)\otimes U\(1\)$, by taking $\alpha_{*}=\alpha_{r}$ or $\alpha_{r-1}$ for the case\\
  & $G=SO\(2r\)$.\\
 { Type IV:}  & $Sp\(r\)/SU\(r\)\otimes U\(1\)$, by taking $\alpha_{*}=\alpha_{r}$  for the case $G=Sp\(r\)$.\\
 { Type V:}  & $E_6/SO\(10\)\otimes U\(1\)$, by taking $\alpha_{*}=\alpha_{1}$ or $\alpha_6$  for the case $G=E_6$.\\
 { Type VI:}  & $E_7/E_6\otimes U\(1\)$, by taking $\alpha_{*}=\alpha_{7}$   for the case $G=E_7$.\\
 \end{tabular} 
 \caption{List of the compact irreducible Hermitian symmetric spaces $G/{\widehat K}\otimes U(1)_{\Lambda}$.}
 \label{listhermitian}
 \end{table}

\section{The construction of harmonic and holomorphic maps}
\label{sec:harmoholo}
\setcounter{equation}{0}

The topological charge relevant for the construction of the self-duality equation on the Hermitian symmetric spaces $G/{\widehat K}\otimes U(1)_{\Lambda}$, described in section \ref{sec:symmetric}, is
\be
Q_{\rm top.}=\frac{i}{32\,\pi}\,\int d^2x\, \varepsilon^{ij}\,\trace\( g\,\Lambda\,g^{-1}\,\sbr{\partial_iX\,X^{-1}}{\partial_jX\,X^{-1}}\)
\lab{topchargelambda}
\ee
where  the integration is over the two-dimensional plane with Cartesian coordinates $x^i$, $i=1,2$,  and $\partial_i$ is the partial derivative w.r.t. $x^i$. We have denoted $\varepsilon^{ij}$ the two-dimensional Levi-Civita symbol, with $\varepsilon^{12}=1$. The overall factor $i$ is introduced to make $Q_{\rm top.}$ real (see below). We work with the orthogonal basis and the normalized trace defined respectively by
\be
{\rm Tr}\(T_a\,T_b\)=\kappa\, \delta_{ab} \;;\qquad\qquad \trace\(T_a\,T_b\)\equiv \frac{1}{\kappa}\,{\rm Tr}\(T_a\,T_b\)= \delta_{ab} \lab{normalizedtr}
\ee
where $T_a$, with $a=1,\,...,\, {\rm dim}\,G$, are the generators of the compact simple Lie algebra of $G$, and  $\kappa$ depending upon the representation where the trace is taken. These generators satisfies $\sbr{T_a}{T_b}=i\,f_{abc}\,T_c $, where $f_{abc}$ is the structure constant. In addition, $\Lambda$ is the generator of the automorphism $\sigma$, defined in \rf{sigmadef}, and $X$ is the so-called principal variable used to parametrize the points of the symmetric space, given by
\be
X\(g\)=g\,\sigma\(g\)^{-1}\,,\qquad\qquad \qquad X\(g\,k\)=X\(g\)\,,\qquad \qquad k\in K\equiv {\widehat K}\otimes U(1)_{\Lambda}
\lab{princvar}
\ee
where $g$ is an element of the compact Lie group $G$ leading to the Hermitian symmetric spaces $G/{\widehat K}\otimes U(1)_{\Lambda}$. Note that $Q_{\rm top.}$ depends only on the fields parameterizing the cosets of $G/{\widehat K}\otimes U(1)_{\Lambda}$, since $g\, k\,\Lambda\,\,\(g\,k\)^{-1}= g\,\Lambda\,g^{-1}$, with $k\in K$, as $\Lambda$ commutes with the elements of the even subgroup under $\sigma$, i.e. $K={\widehat K}\otimes U(1)_{\Lambda}$.

In the appendix \ref{sec:proofq} we show that $Q_{\rm top.}$ is invariant under any smooth (homotopic) variations of the fields parameterizing the Hermitian symmetric space $G/{\widehat K}\otimes U(1)_{\Lambda}$, as long as such fields  go to  constant values at infinity on the plane $x^1\,x^2$.  Therefore,  for topological considerations, one can identify the infinity on the plane to a point, and the integration in \rf{topchargelambda} is in fact over such a two-sphere $S^2$.  The topological charge $Q_{\rm top.}$ will play the role of \rf{topcharge} in the construction of the self-duality equations. 

We now project $g^{-1}\,\partial_ig$ into the even and odd subspaces of the automorphism $\sigma$ as in \rf{splittingcalg} and \rf{piplusminusdef} 
\be
P_i=\frac{1}{2}\(g^{-1}\,\partial_ig-\sigma\(g^{-1}\,\partial_ig\)\)\qquad\qquad\qquad
K_i=\frac{1}{2}\(g^{-1}\,\partial_ig+\sigma\(g^{-1}\,\partial_ig\)\)
 \lab{ws}
\ee
On the other hand, we can use \rf{hermitianniceprop} to split $P_i$ into the $\pm 1$ subspaces, i.e. 
\be
P_i \equiv  P_i^{(+)}+P_i^{(-)}\,,\qquad\qquad\quad\qquad \sbr{\Lambda}{P_i^{(\pm)}}=\pm P_i^{(\pm)}
\lab{piplusminusdef} 
\ee
Using \rf{princvar} and \rf{ws} we can write 
\be
\partial_iX\,X^{-1}= 2\,g\, P_i\, g^{-1}
\lab{dxpdef}
\ee
Consequently, $Q_{\rm top.}$ becomes
\be
Q_{\rm top.}=\frac{i}{8\,\pi}\,\int d^2x\, \varepsilon^{ij}\,\trace\( \sbr{\Lambda}{P_i}\,P_j\)=\frac{i}{4\,\pi}\,\int d^2x\, \varepsilon^{ij}\,\trace \(P_i^{(+)}\,P_j^{(-)}\)
\lab{topchargelambda2}
\ee
where we have used the fact that, as a consequence of \rf{killingform} and \rf{granding2}, 
\be
{\rm Tr}\(P_i^{(+)}\,P_j^{(+)}\)={\rm Tr}\(P_i^{(-)}\,P_j^{(-)}\)=0
\lab{zerotrace}
\ee
Since $G$ is compact, any representation of it is equivalent to a unitary one. Then the elements $g$ of $G$ are unitary, i.e $g^{\dagger}=g^{-1}$, and so is the principal variable, i.e. $X^{\dagger}=X^{-1}$. Since $\Lambda$ is hermitian, it follows that $Q_{\rm top.}$, defined in \rf{topchargelambda}, is indeed real. From \rf{ws} we then get that $P_i$ is anti-hermitian, i.e. $P_i^{\dagger}=-P_i$. Writing 
\be
P_i^{(\pm)}= \sum_{\kappa=1}^{{\rm dim} {\cal P}/2} \sqrt{\frac{\alpha_{\kappa}^2}{2}}\,P_i^{(\pm \, ,\, \kappa)} \, E_{\pm \alpha_{\kappa}}
\lab{componentspi}
\ee 
we get that
\be
\(P_i^{(+ \, ,\, \kappa)}\)^*=- P_i^{(- \, ,\, \kappa)}
\lab{hermiticitypi}
\ee
where we have used the hermiticity condition  $E_{\alpha_{\kappa}}^{\dagger}=E_{-\alpha_{\kappa}}$ (see paragraph below \rf{hermitianniceprop}). Therefore, using the Killing form \rf{killingform} we get that
\br
Q_{\rm top.}&=&\frac{i}{8\,\pi\,\kappa}\,\int d^2x\, \sum_{\kappa=1}^{{\rm dim} {\cal P}/2} \varepsilon^{ij}\,\left[P_i^{(+\,,\,\kappa)}\,P_j^{(-\,,\,\kappa)}-P_i^{(-\,,\,\kappa)}\,P_j^{(+\,,\,\kappa)}\right]
\nonumber\\
&=&
\frac{1}{8\,\pi\,\kappa}\,\int d^2x\, \sum_{\kappa=1}^{{\rm dim} {\cal P}/2} \left[ P_i^{(+\,,\,\kappa)}\,\(i\,\varepsilon^{ij}\,P_j^{(+\,,\,\kappa)}\)^*
+\(P_i^{(+\,,\,\kappa)}\)^*\,i\,\varepsilon^{ij}\,P_j^{(+\,,\,\kappa)}
\right]
\lab{topchargelambda3}
\er

We shall take the energy functional as
\be
E=-\frac{1}{e_0^2}\,\int d^2x\,\trace\(\partial_i X\,X^{-1}\)^2
\lab{energydef}
\ee
where $e_0$ is some coupling constant, and $X$ is the principal variable defined in \rf{princvar}. Note that $E$, given in \rf{energydef}, is real and positive (since $\partial_i X\,X^{-1}$ is anti-hermitian), and it corresponds, in fact, to the static kinetic energy of the non-linear sigma model defined on the Hermitian symmetric space $G/{\widehat K}\otimes U(1)_{\Lambda}$. In fact, the distance on such symmetric space can be described as $ds^2=-{\rm Tr}\(d X\,X^{-1}\)^2$, with $d$ being the exterior derivative on $G/{\widehat K}\otimes U(1)_{\Lambda}$. The (static) Euler-Lagrange equations associated to the energy functional \rf{energydef} are
\be
\partial^i\(\partial_i X\,X^{-1}\)=0
\lab{eulerlagsymmspace}
\ee

Using \rf{dxpdef}, \rf{piplusminusdef}, \rf{zerotrace}\, \rf{componentspi} and \rf{hermiticitypi}, one can write \rf{energydef} as
\be
E=\frac{4}{e_0^2\,\kappa}\,\int d^2x\,\sum_{\kappa=1}^{{\rm dim} {\cal P}/2}\left[P_i^{(+\,,\,\kappa)}\(P_i^{(+\,,\,\kappa)}\)^*+i\,\varepsilon^{ij}\,P_j^{(+\,,\,\kappa)}\(i\,\varepsilon^{ik}\,P_k^{(+\,,\,\kappa)}\)^*
\right]
\lab{energydef2}
\ee

Comparing \rf{topchargelambda3} with \rf{topcharge}, and \rf{energydef2} with \rf{energy}, we are led to make the identifications (disregarding overall factors) 
\be
{\cal A}_{\alpha} \rightarrow P_i^{(+\,,\,\kappa)}\,, \qquad\qquad {\rm} \qquad\qquad {\widetilde{\cal A}}_{\alpha}\rightarrow i\,\varepsilon^{ij}\,P_j^{(+\,,\,\kappa)}
\lab{identify}
\ee
and so, the self-duality equations \rf{sdeqs} becomes
\be
P_i^{(+\,,\,\kappa)}=\pm i\,\varepsilon_{ij}\,P_j^{(+\,,\,\kappa)}
\lab{seldualeqsfinal}
\ee

Therefore, the topological charge \rf{topchargelambda3} and the energy \rf{energydef2} fit into the scheme of the generalized self-duality construction, described in section 
\ref{sec:self-duality}. Consequently, solutions of the self-duality equations \rf{seldualeqsfinal} are solutions of the Euler-Lagrange equations \rf{eulerlagsymmspace}, associated to the energy functional \rf{energydef2}. In addition, such self-dual solutions saturate the lower bound \rf{bound}. Indeed, we can write \rf{energydef2} as
\br
E&=&\frac{4}{e_0^2\,\kappa}\,\int d^2x\,\sum_{\kappa=1}^{{\rm dim} {\cal P}/2} \left[P_i^{(+\,,\,\kappa)}\mp i\,\varepsilon^{ij}\,P_j^{(+\,,\,\kappa)}
\right]\left[\(P_i^{(+\,,\,\kappa)}\)^*\mp \(i\,\varepsilon^{ik}\,P_k^{(+\,,\,\kappa)}\)^*\right]
\ \nonumber\\
&\pm & \frac{4}{e_0^2\,\kappa}\,\int d^2x\,\sum_{\kappa=1}^{{\rm dim} {\cal P}/2} \left[ P_i^{(+\,,\,\kappa)}\,\(i\,\varepsilon^{ij}\,P_j^{(+\,,\,\kappa)}\)^*
+\(P_i^{(+\,,\,\kappa)}\)^*\,i\,\varepsilon^{ij}\,P_j^{(+\,,\,\kappa)}
\right] 
\lab{energydef3}
\er
Note that that the upper sign in \rf{seldualeqsfinal} imply that the topological charge is positive, and the lower sign that it is negative. Indeed, 
\br
P_i^{(+\,,\,\kappa)}= i\,\varepsilon_{ij}\,P_j^{(+\,,\,\kappa)}\qquad\qquad &\rightarrow& \qquad\qquad Q_{\rm top.}>0
\nonumber\\
P_i^{(+\,,\,\kappa)}= -i\,\varepsilon_{ij}\,P_j^{(+\,,\,\kappa)}\qquad\qquad &\rightarrow& \qquad\qquad Q_{\rm top.}<0 \lab{selfdualP}
\er
Therefore, since the first line in \rf{energydef3} is positive, we have the lower bound on the energy
\be
E\geq \frac{32\,\pi}{e_0^2}\, \mid Q_{\rm top.}\mid
\ee
Such a bound is saturated by the solutions of the self-duality equations \rf{seldualeqsfinal}. 

From \rf{killingform}, \rf{componentspi}, and \rf{piplusminusdef} we have that
\be
P_i^{(+\,,\,\kappa)}=\sqrt{\frac{\alpha_{\kappa}^2}{2}}\, {\rm Tr}\(P_i^{(+)}\,E_{-\alpha_{\kappa}}\)
=\sqrt{\frac{\alpha_{\kappa}^2}{2}}\, {\rm Tr}\(P_i\,E_{-\alpha_{\kappa}}\)
=\sqrt{\frac{\alpha_{\kappa}^2}{2}}\, {\rm Tr}\(g^{-1}\,\partial_i g\,E_{-\alpha_{\kappa}}\)
\lab{piplusmcform}
\ee
where we use \rf{killingform}, the fact that ${\rm Tr}\(P_i^{(-)}\,E_{-\alpha_{\kappa}}\)=0$, and that the Killing form is invariant under the automorphism $\sigma$. Replacing  \rf{piplusmcform} into \rf{seldualeqsfinal} we observe that the self-duality equations constitute the Cauchy-Riemann equations for the components ${\rm Tr}\(g^{-1}\,\partial_i g\,E_{-\alpha_{\kappa}}\)$ of the Maurer-Cartan form $g^{-1}\,\partial_i g$. Indeed, introducing the complex coordinates
\be
z\equiv x_1-i\,x_2\,, \qquad {\bar z}\equiv x_1+i\,x_2\,,\qquad \partial_1=\partial_z+\partial_{{\bar z}}\,, \qquad \partial_2=-i\,\(\partial_z-\partial_{{\bar z}}\) \lab{zcoordinates}
\ee
one gets that the upper sign in \rf{seldualeqsfinal} imply that
\be
{\rm Tr}\(g^{-1}\,\partial_{\bar z} g\,E_{-\alpha_{\kappa}}\)=0\qquad\mbox{\rm and so}\qquad 
{\rm Tr}\(g^{-1}\,\partial_z g\,E_{\alpha_{\kappa}}\)=0
\qquad \qquad \mbox{\rm for any $\alpha_{\kappa}$}
\lab{holomc}
\ee
or equivalently $P_{\bar{z}}^{(+)} =P_{{z}}^{(-)} =0$. 
The lower sign in \rf{seldualeqsfinal} imply that
\be
{\rm Tr}\(g^{-1}\,\partial_z g\,E_{-\alpha_{\kappa}}\)=0\qquad\mbox{\rm and so}\qquad
{\rm Tr}\(g^{-1}\,\partial_{\bar z} g\,E_{\alpha_{\kappa}}\)=0
\qquad\qquad \mbox{\rm for any $\alpha_{\kappa}$}
\lab{antiholomc}
\ee
or equivalently $P_{{z}}^{(+)} =P_{\bar{z}}^{(-)} =0 $, where we have used the fact $E_{\alpha_{\kappa}}^{\dagger}=E_{-\alpha_{\kappa}}$, and that $g$ is unitary, and so $\(g^{-1}\,\partial_i g\)^{\dagger}=-g^{-1}\,\partial_i g$. From \rf{hermiticitypi} and \rf{piplusmcform}, we have that
\be
P_i^{(-\,,\,\kappa)}=-\sqrt{\frac{\alpha_{\kappa}^2}{2}}\, \({\rm Tr}\(g^{-1}\,\partial_i g\,E_{-\alpha_{\kappa}}\)\)^*=\sqrt{\frac{\alpha_{\kappa}^2}{2}}\, {\rm Tr}\(g^{-1}\,\partial_i g\,E_{\alpha_{\kappa}}\)
\lab{pminuskappacomplex}
\ee
Note that in \rf{hermiticitypi}, and also in \rf{pminuskappacomplex}, we have assumed that the index $i$ corresponds to the real Cartesian coordinates $x^i$, $i=1,2$, and not to the complex coordinates $z$ and ${\bar z}$. 

Therefore, from \rf{dxpdef}, \rf{piplusminusdef}, \rf{componentspi}, \rf{piplusmcform} and \rf{pminuskappacomplex} we get
\be
\partial_iX\,X^{-1}= 2\,\sum_{\alpha_{\kappa}>0} \frac{\alpha_{\kappa}^2}{2}\, g\, \left[ {\rm Tr}\(g^{-1}\,\partial_i g\,E_{-\alpha_{\kappa}}\)\,E_{\alpha_{\kappa}}+ {\rm Tr}\( g^{-1}\,\partial_i g\,E_{\alpha_{\kappa}}\)\,E_{-\alpha_{\kappa}}\right]\, g^{-1}
\ee
From \rf{holomc} we get that the upper sign in \rf{seldualeqsfinal} imply that
\be
{\rm Tr}\left[g^{-1}\,\(\partial_{\bar z}X\)\,\sigma\(g\)\,E_{-\alpha_{\kappa}}\right]={\rm Tr}\left[\partial_{\bar z}X\,X^{-1}\,g\,E_{-\alpha_{\kappa}}g^{-1}\right]=0
\lab{holox}
\ee
and from \rf{antiholomc} we get that the lower sign in \rf{seldualeqsfinal} imply that
\be
{\rm Tr}\left[g^{-1}\,\(\partial_{z}X\)\,\sigma\(g\)\,E_{\alpha_{-\kappa}}\right]={\rm Tr}\left[\partial_z X\,X^{-1}\,g\,E_{-\alpha_{\kappa}}g^{-1}\right]=0
\lab{antiholox}
\ee
Therefore, the self-duality equations \rf{seldualeqsfinal} imply that the principal variable $X$, defined in \rf{princvar}, have to be either holomorphic or anti-holomorphic, in the sense of \rf{holox} and \rf{antiholox} respectively. \\

Summarizing:
\begin{enumerate}
\item We have shown that the quantity $Q_{\rm top.}$, defined in \rf{topchargelambda}, is homotopicaly invariant if the principal variable $X\(g\)$, defined in \rf{princvar}, goes to a constant at infinity on  the plane $x^1\,x^2$ (see \rf{bcforx} in appendix \ref{sec:proofq}). With such a boundary condition the plane can be considered, for topological considerations, as the two-sphere $S^2$. Therefore, $X\(g\)$ provides a map from  $S^2$  to the Hermitian symmetric space $G/{\widehat K}\otimes U(1)_{\Lambda}$. 
\item The maps $X\(g\)$ that satisfy the self-duality equations \rf{seldualeqsfinal} are harmonic as they minimize the energy \rf{energydef}, and satisfy the Euler-Lagrange equations \rf{eulerlagsymmspace} associated to it. 
\item In addition, the  maps $X\(g\)$ that satisfy the self-duality equations \rf{seldualeqsfinal} have to be holomorphic or anti-holomorphic, in the sense of \rf{holox} and \rf{antiholox} respectively. 
\end{enumerate}

Therefore, the maps $X\(g\)$, from $S^2 \rightarrow  G/{\widehat K}\otimes U(1)_{\Lambda}$, are harmonic and holomorphic or anti-holomorphic, in the sense of \rf{holox} and \rf{antiholox} respectively. 

\section{The rational map ansatz in three dimensions}
\label{sec:rationalmap}
\setcounter{equation}{0}

We now show how to construct the rational map ansatz in three spatial dimensions for any compact simple Lie group $G$ that leads to an Hermitian symmetric space, i.e. those $G$ appearing in the Table \ref{listhermitian}.  It is known that the third homotopy group of a simple compact Lie group $G$ is the integers, i.e. $\pi_3\(G\)= \IZ$. The topological charge associated to such homotopy group is given by
\be
 Q= \frac{i}{48\,\pi^2}\int d^3y\; \ve_{ijk}\,\trace\(R_i\,R_j\,R_k\)
 \lab{topcharge2}
 \ee
 where $y_i$, $i=1,2,3$, are the Cartesian coordinates in $\IR^3$, and $R_i$ being the Maurer-Cartan one-form, i.e. 
 \be
R_{i} \equiv i\,\partial_{i}U\,U^{-1}
\lab{rdef}
\ee
with $U$ being an element of the group $G$. The regular maps are those where $U$ goes to a constant at spatial infinity, and so for topological reasons we can compactify $\IR^3$ into the three sphere $S^3$.

In order to construct the rational map ansatz we foliate $\IR^3$ with two spheres $S^2$ centered at the origin, and instead of using spherical polar coordinates, we stereographic project each $S^2$ onto a plane $\IR^2$, and we parameterize that plane by a complex coordinate $z$. So, we use the coordinate system $\(r\,,\,z\,,\,{\bar z}\)$ defined by ($z=z_1+iz_2$)
\be
y_1= r\, \frac{i\(\bar{z}-z\)}{1+\mid z\mid^2} \; ; \qquad 
y_2= r\, \frac{z+\bar{z}}{1+\mid z\mid^2} \; ; \qquad 
y_3= r\, \frac{\(-1+\mid z\mid^2\)}{1+\mid z\mid^2} 
\lab{coordinates}
\ee
and the Euclidean metric on $\IR^3$ becomes 
\be
ds^2=dr^2+ \frac{4\,r^2}{\(1+\mid z\mid^2\)^2} \, dz\,d{\bar z} \lab{metric}
\ee

We define the rational map ansatz in three dimensions as  
\be
U= g\, e^{i\,f\(r\)\, \Lambda}\, g^{-1}=e^{i\,f\(r\)\,g\,\Lambda\,g^{-1}}
\lab{holog}
\ee
where $g$ is  an element of the compact Lie group $G$, and it  depends only $z$ and ${\bar z}$, i.e. $g=g\(z\,,\,{\bar z}\)$. $\Lambda$ is defined in \rf{sigmadef}, and $f\(r\)$ is a radial profile function.  Note that the term $g\,\Lambda\,g^{-1}$ depends only on the fields parameterizing the cosets in $G/{\widehat K}\otimes U(1)_{\Lambda}$, since $g\, k\,\Lambda\,\,\(g\,k\)^{-1}= g\,\Lambda\,g^{-1}$, with $k\in K$, as $\Lambda$ commutes with the elements of the even subgroup under $\sigma$, i.e. $K={\widehat K}\otimes U(1)_{\Lambda}$. In fact, $g$ can be taken as a principal variable \rf{princvar}, i.e. $g\(w\)=w\,\sigma\(w\)^{-1}$, with $w$ an element of $G$. In such a case, we have that $\sigma\(g\)=g^{-1}$, and so $X\(g\)=g\,\sigma\(g\)^{-1}=g^2$. In addition, as we have shown in section \ref{sec:harmoholo}, the fields $g$, or equivalently $X\(g\)$, define harmonic and holomorphic (or anti-holomorphic)  maps from the two sphere $S^2$ to the Hermitian symmetric spaces $G/{\widehat K}\otimes U(1)_{\Lambda}$. We give a concrete construction of the element $g$ in section \ref{sec:constructg}.

It then follows that
 \be
R_{i} = i\,V^{-1}\left[i\,\partial_if\,\Lambda
-e^{i\,f\,\Lambda/2}\,\,g^{-1}\partial_ig\,e^{-i\,f\,\Lambda/2}+e^{-i\,f\,\Lambda/2}\,g^{-1}\, \partial_ig\,e^{i\,f\,\Lambda/2}\right]\,V
\ee
with $V=e^{-i\,f\,\Lambda/2}\,g^{-1}$. Using \rf{ws} and the fact that $K_i$ commutes with $\Lambda$, and so it drops from that expression, we get
 \be
R_{i} = i\,V^{-1}\left[i\,\partial_if\,\Lambda
-e^{i\,f\,\Lambda/2}\, P_i\,e^{-i\,f\,\Lambda/2}+e^{-i\,f\,\Lambda/2}\, P_i\,e^{i\,f\,\Lambda/2}\right]\,V
\ee
Since we are dealing with Hermitian symmetric spaces we can  split $P_i$ into the $\pm 1$ subspaces, as we did in \rf{piplusminusdef}. We then get that  
 \be
R_{i} = -V^{-1}\, \Sigma_i\, V \lab{rv}
\ee
with 
\br
\Sigma_i&\equiv &\partial_if\,\Lambda
- 2\,\sin \frac{f}{2}\(P^{(+)}_i-P^{(-)}_i\)
\lab{Sigmadef}
\er
Therefore
\be
 \ve_{ijk}\,\trace\(R_i\,R_j\,R_k\)=-\frac{1}{2}\,\ve_{ijk}\,\trace\(\Sigma_i\,\sbr{\Sigma_j}{\Sigma_k}\)
 =12\,\ve_{ijk}\,\partial_if\,\sin^2\frac{f}{2}\,\trace\(P^{(+)}_j\,P^{(-)}_k\) \lab{pred}
\ee
and so
\br
 Q&=& \frac{i}{4\,\pi^2}\int dr\,dz\,d{\bar z}\;\partial_rf\,\sin^2\frac{f}{2}\,\trace\(P^{(+)}_z\,P^{(-)}_{{\bar z}}-P^{(+)}_{{\bar z}}\,P^{(-)}_z\)
 \nonumber\\
 &=&\frac{1}{2\,\pi}\,\left[f\(r\)-\sin f\(r\)\right]_{r=0}^{r=\infty}\;\frac{i}{4\,\pi}\,
 \int dz\,d{\bar z}\,\trace\(P^{(+)}_z\,P^{(-)}_{{\bar z}}-P^{(+)}_{{\bar z}}\,P^{(-)}_z\)
 \lab{pretopcharge3d}
\er
Comparing \rf{pretopcharge3d}  with \rf{topchargelambda2} we get that
\be
Q=\frac{1}{2\,\pi}\,\left[f\(r\)-\sin f\(r\)\right]_{r=0}^{r=\infty}\; Q_{\rm top.}
\lab{charge3d}
\ee
So, the three dimensional topological charge \rf{topcharge2} factors into the product of the two dimensional topological charge \rf{topchargelambda}, built out of the harmonic and (anti)holomorphic self-dual maps from $S^2$ to the Hermitian symmetric spaces, and the boundary conditions on the profile function $f\(r\)$.   

\subsection{The construction of the element $g$ in the ansatz \rf{holog}}
\label{sec:constructg}

We want the element $g$ appearing in the ansatz \rf{holog} to parameterize the hermitian symmetric spaces $G/{\widehat K}\otimes U(1)_{\Lambda}$, and to depend only on the angular variables $z$ and ${\bar z}$, defined in  \rf{coordinates}. So, it will provide the harmonic and holomorphic maps from $S^2$ to $G/{\widehat K}\otimes U(1)_{\Lambda}$. 

The key ingredient for the construction of $g$ is a Lie algebra element $S$ lying on the subspace generated by the step operators $E_{\alpha_{\kappa}}$ associated to the roots $\alpha_{\kappa}$ containing the simple root $\alpha_*$ in their expasion. See \rf{alphastardef} and \rf{evenoddgenerators} for details. So, we have
\be
S=\sum_{\kappa} w_{\kappa}\,E_{\alpha_{\kappa}}\;; \qquad\qquad \sbr{\Lambda}{S}=S
\lab{sdef}
\ee
where $w_{\kappa}$ are functionals of the fields (holomorphic or anti-holomorphic) parameterizing the hermitian symmetric spaces $G/{\widehat K}\otimes U(1)_{\Lambda}$. Therefore, we have that
\be
S^{\dagger}=\sum_{\kappa} w_{\kappa}^*\,E_{-\alpha_{\kappa}}\;; \qquad\qquad \sbr{\Lambda}{S^{\dagger}}=-S^{\dagger}
\lab{sdaggerdef}
\ee
We want to find representations of the Lie algebra of $G$ where the matrix associated to $S$ is nilpotent, and an eigenvector of the hermitian matrix $S\,S^{\dagger}$, which we impose to have non-vanishing trace, i.e.
\be
S^2=0\;;\qquad\qquad \qquad \qquad\(S\,S^{\dagger}\)\,S=\omega\,S
\lab{omegadef}
\ee
with the eigenvalue $\omega$ being real and  non-negative. Consequently we also have that
\be
{S^{\dagger}}^2=0\;;\qquad\qquad \qquad \qquad\(S^{\dagger}\,S\)\,S^{\dagger}=\omega\,S^{\dagger}
\lab{omegadef2}
\ee
The element $g$ is chosen to be of the form
\be
g\equiv e^{i\,S}\,e^{\varphi\,\sbr{S}{S^{\dagger}}}\,e^{i\,S^{\dagger}}
\lab{gdef}
\ee
for a real function $\varphi$ to be determined. From \rf{sigmadef}, \rf{sdef} and \rf{sdaggerdef} we have that
\be
\sigma\(g\)=e^{-i\,S}\,e^{\varphi\,\sbr{S}{S^{\dagger}}}\,e^{-i\,S^{\dagger}}=g^{\dagger}
\lab{sigmag}
\ee
In order for $g$ to be an element of the compact simple Lie group $G$, it has to be unitary, i.e. $g^{\dagger}=g^{-1}$, and so 
\be
\sigma\(g\)=g^{-1}
\lab{sigmaggminus1}
\ee
As we will see, the unitarity condition will determine the real function $\varphi$. Using \rf{omegadef} and \rf{omegadef2} we have that
\be
\left[S,\,S^\dagger \right]^{n}= \omega^{n-1}\, \(S\,S^\dagger +\(-1\)^n S^\dagger\,S \) 
\lab{commssdagger}
\ee
Therefore, for the case $n=2$, we conclude that $\omega$ is determined in terms of $S$ and $S^{\dagger}$ as
\be 
\omega = \frac{\Tr \(\left[S,\,S^\dagger \right]^2\)}{2\,\Tr \(S\,S^\dagger\)}
\ee 
Using \rf{commssdagger} we get that
\br
 e^{\pm \varphi\,\left[S,\,S^\dagger \right]} & = &  \one +\frac{1}{\omega}\,\sum_{n=1}\,\frac{\(\pm \varphi \,\omega\)^{n}}{n!}\, \(S\,S^\dagger +\(-1\)^n S^\dagger\,S \)\nonumber\\
 & = &  \one +\frac{1}{\omega}\, \left[\(e^{\pm \varphi \,\omega}-1\) \,S\,S^\dagger + \(e^{\mp \varphi \,\omega}-1\) \,S^\dagger\,S \right]
 \lab{expcomms}
\er
Therefore, from \rf{omegadef}, \rf{omegadef2} and \rf{expcomms} we have
\br
e^{\pm \varphi\,\left[S,\,S^\dagger \right]} \, S & = & e^{\pm \varphi \,\omega}\,S\;;\qquad\qquad\qquad
 S \,e^{\pm \varphi\,\left[S,\,S^\dagger \right]} =  e^{\mp \varphi \,\omega}\,S
 \nonumber\\
 e^{\pm \varphi\,\left[S,\,S^\dagger \right]} \, S^{\dagger} & = & e^{\mp \varphi \,\omega}\,S^{\dagger}\;;\qquad\qquad\quad\;
 S^{\dagger} \,e^{\pm \varphi\,\left[S,\,S^\dagger \right]} =  e^{\pm \varphi \,\omega}\,S^{\dagger}
 \lab{niceeigenvec}
\er
or equivalently
\br
e^{ \varphi\,\left[S,\,S^\dagger \right]} \, S & = & S\,e^{ -\varphi\,\left[S,\,S^\dagger \right]} \;;\qquad\qquad\qquad
 S \,e^{ \varphi\,\left[S,\,S^\dagger \right]} =  e^{ -\varphi\,\left[S,\,S^\dagger \right]}\,S
 \nonumber\\
 e^{ \varphi\,\left[S,\,S^\dagger \right]} \, S^{\dagger} & = & S^{\dagger}\,e^{- \varphi\,\left[S,\,S^\dagger \right]} \;;\qquad\qquad\quad\;
 S^{\dagger} \,e^{ \varphi\,\left[S,\,S^\dagger \right]} =  e^{- \varphi\,\left[S,\,S^\dagger \right]}\,S^{\dagger}
\er

The relations \rf{omegadef}, \rf{omegadef2} and \rf{niceeigenvec} imply that the element $g$, given \rf{gdef}, satisfy
\br
g^{\dagger}&=&\(\one -i\,S\)e^{ \varphi\,\left[S,\,S^\dagger \right]}\(\one-i\,S^{\dagger}\)
\nonumber\\
&=& e^{ \varphi\,\left[S,\,S^\dagger \right]} -e^{- \varphi \,\omega}\,S\,S^{\dagger}-i\,e^{- \varphi \,\omega}\,S-i\,e^{- \varphi \,\omega}\,S^{\dagger}
\er
and
\br
g^{-1}&=&\(\one -i\,S^{\dagger}\)e^{- \varphi\,\left[S,\,S^\dagger \right]}\(\one-i\,S\)
\nonumber\\
&=& e^{ -\varphi\,\left[S,\,S^\dagger \right]} -e^{-\varphi \,\omega}\,S^{\dagger}\,S-i\,e^{- \varphi \,\omega}\,S-i\,e^{- \varphi \,\omega}\,S^{\dagger}
\er
Therefore
\be
g^{\dagger}=g^{-1}\quad\rightarrow\qquad e^{ \varphi\,\left[S,\,S^\dagger \right]} -e^{- \varphi \,\omega}\,S\,S^{\dagger}=e^{ -\varphi\,\left[S,\,S^\dagger \right]} -e^{-\varphi \,\omega}\,S^{\dagger}\,S
\lab{gunitary}
\ee
Multiplying both sides of \rf{gunitary} by $S$ from the left, we get that
\be
\left[\(1+\omega\)\,e^{-\varphi \,\omega}-e^{\varphi \,\omega}\right]\,S=0
\ee
and so
\be
\varphi=\frac{\ln\sqrt{1+\omega}}{\omega}
\lab{phidef}
\ee
That concludes the construction of the matrix $g$, showing that it is unitary and satisfy \rf{sigmaggminus1}.\\ 

Using \rf{expcomms}  and \rf{niceeigenvec} we get that the group element $g$, given in \rf{gdef}, for 
$\varphi$ given in \rf{phidef}, can be written as 
\be
g =   \one +\frac{1}{\vartheta}\, \left[i\(S+S^\dagger\) -\frac{1}{\vartheta+1} \,\left(S\,S^\dagger+S^\dagger\,S\right) \right] ; \qquad  \qquad\vartheta\equiv \sqrt{1+\omega}
 \lab{gdef2}
\ee
Note that, using \rf{omegadef} and \rf{omegadef2} we get that
\be
e^{i\,\frac{\theta}{\sqrt{\omega}}\(S+S^{\dagger}\)}=\one+i\,\frac{\sin\theta}{\sqrt{\omega}}\,\(S+S^{\dagger}\)+\frac{\(\cos\theta -1\)}{\omega}\,\(S\,S^{\dagger} + S^{\dagger}\,S\)
\ee
Therefore, $g$ given in \rf{gdef2} (or \rf{gdef}), can be written as 
\be 
g = e^{i\,\frac{\theta}{\sqrt{\omega}}\,\(S+S^\dagger\)}\qquad\qquad {\rm with}\qquad\qquad \theta= {\rm arcsin}\(\frac{\sqrt{\omega}}{\sqrt{1+\omega}}\)
\lab{gexp}
\ee

From \rf{finerstructure} we have that the abelian subspaces generated by $E_{\alpha_{\kappa}}$ and $E_{-\alpha_{\kappa}}$, constitute representations of the zero grade subgroup which Lie algebra is ${\cal G}_0={\cal K}$. Therefore,
\be
e^{\varphi\,\sbr{S}{S^{\dagger}}}\, E_{-\alpha_{\kappa}}\, e^{-\varphi\,\sbr{S}{S^{\dagger}}}=E_{-\alpha_{\kappa^{\prime}}}\, R_{\kappa^{\prime}\, \kappa}\(e^{\varphi\,\sbr{S}{S^{\dagger}}}\)
\ee
where $R\(e^{\varphi\,\sbr{S}{S^{\dagger}}}\)$ is the matrix of the group element $e^{\varphi\,\sbr{S}{S^{\dagger}}}$ in that represention. 

Consequently, from \rf{holomc}, \rf{gdef},   \rf{finerstructure} and \rf{granding2} we get 
\br
0&=&{\rm Tr}\(g^{-1}\partial_{\bar z} g\, E_{-\alpha_{\kappa}}\)
\nonumber\\
&=&{\rm Tr}\left[\( e^{-\varphi\,\sbr{S}{S^{\dagger}}}\, i\,\partial_{\bar z}S\,e^{\varphi\,\sbr{S}{S^{\dagger}}}\,+ e^{-\varphi\,\sbr{S}{S^{\dagger}}}\, \partial_{\bar z} e^{\varphi\,\sbr{S}{S^{\dagger}}}\, 
+  i\partial_{\bar z}S^{\dagger}\) e^{i\,S^{\dagger}}\,E_{-\alpha_{\kappa}}\,e^{-i\,S^{\dagger}} \,\right]
\nonumber\\
&=&{\rm Tr}\left[ i\,\partial_{\bar z}S\;e^{\varphi\,\sbr{S}{S^{\dagger}}}\, E_{-\alpha_{\kappa}}\,e^{-\varphi\,\sbr{S}{S^{\dagger}}}\right]
\nonumber\\
&=&{\rm Tr}\left[ i\,\partial_{\bar z}S\,E_{-\alpha_{\kappa^{\prime}}}\right]\, R_{\kappa^{\prime}\, \kappa}\(e^{\varphi\,\sbr{S}{S^{\dagger}}}\)
\er
Since $R\(e^{\varphi\,\sbr{S}{S^{\dagger}}}\)$ is invertible, we get that ${\rm Tr}\left[ i\,\partial_{\bar z}S\,E_{-\alpha_{\kappa^{\prime}}}\right]=0$ for any $\alpha_{\kappa^{\prime}}$. 
Therefore, we get that $S$ is holomorphic
\be
\partial_{{\bar z}} S=0
\lab{finalholos}
\ee
Performing a similar calculation starting from \rf{antiholomc} we see that  $S$ must be anti-holomorphic
\be
\partial_{z} S=0
\lab{finalantiholos}
\ee
Consequently, the upper sign in \rf{seldualeqsfinal}, or equivalent \rf{holomc}, imply that $S$ must be holomorphic. The lower sign in \rf{seldualeqsfinal}, or equivalent \rf{antiholomc}, imply that $S$ must be anti-holomorphic.  Conversely, \rf{finalholos} (or \rf{finalantiholos}) imply the self-duality  \rf{seldualeqsfinal} with the upper sign (or lower sign), and so it leads to self-dual solutions of the non-linear sigma model on the Hermitian symmetric space  $G/{\widehat K}\otimes U(1)_{\Lambda}$.

As we will see in the examples of section \ref{sec:examples}, the holomorphic (or anti-holomorphic) fields appearing on the entries of the matriz $S$ shall be taken to be ratios of polynomials on the complex variable $z$ (or ${\bar z}$), and that is what leads to the rational character of the  map \cite{mantonbook}. \\

That completes the parameterization of the group element $U$ in the rational ansatz \rf{holog}. \\

The argument above shows the equivalence of $\partial_{\bar{\chi}} S=0 \Leftrightarrow P_{\bar{\chi}}^{(+)}=0$, for $\chi=z,\,\bar{z}$. Then, the holomorphic and anti-holomorphic ansatz can be written in a more compact way as
\be
\pa_{\bar{\chi}} S=\pa_{{\chi}} S^\dagger=0\quad \Leftrightarrow\quad  P_{\bar{\chi}}^{(+)}=P_{{\chi}}^{(-)}=0 \quad\qquad  \mbox{with} \quad\qquad \chi\equiv \left\{\begin{array}{ll} z, & S=S(z)\\ \bar{z}, & S=S(\bar{z})\end{array} \right. \lab{wgp}
\ee
This equivalence can also be proven from the explicit calculation of $P_i^{(\pm)}$ using \rf{ws}, \rf{piplusminusdef}, and \rf{gdef2}. In fact, we can find a Hermitian and invertible matrix $ \Omega \equiv \one + \frac{S\,S^\dagger+ S^\dagger\,S}{1+\vartheta}$, where $\Omega^{-1} =\one - \frac{S\,S^\dagger+S^\dagger\,S}{\vartheta\,\(1+\vartheta\)} $, that leads to $\Omega\,P_i^{(+)}\,\Omega = i\,\pa_i S$, which clearly implies  \rf{wgp}. In addition, these calculations, presented in Appendix \ref{sec:appproof}, also show that the even operators and the non-vanishing components of the odd operators defined in \rf{ws} and \rf{piplusminusdef} reduce to
\br
P_\chi^{(+)} &=& i\,\frac{\(1+\vartheta\)^2}{\vartheta}\,\pa_\chi \(\frac{S}{\(1+\vartheta\)^2}\);\qquad\quad P_{\bar{\chi}}^{(-)}= i\,\frac{\(1+\vartheta\)^2}{\vartheta}\,\pa_{\bar{\chi}} \(\frac{S^\dagger}{\(1+\vartheta\)^2}\) \lab{pir}\\
K_\chi & = & + \frac{1}{\vartheta\,\(1+\vartheta\)}\,\left[\frac{1}{2}\,\frac{\pa_\chi \omega}{\vartheta\(1+\vartheta\)}\,\left[S ,\, S^\dagger\right] - \pa_\chi \left[S ,\, S^\dagger\right]\right]   \\
K_{\bar{\chi}} & = & - \frac{1}{\vartheta\,\(1+\vartheta\)}\,\left[\frac{1}{2}\,\frac{\pa_{\bar{\chi}} \omega}{\vartheta\(1+\vartheta\)}\,\left[S ,\, S^\dagger\right] - \pa_{\bar{\chi}} \left[S ,\, S^\dagger\right]\right]  
\er

The relations \rf{wgp} allow us to write the components of \rf{Sigmadef} in the coordinates \rf{metric}  as 
\br
\Sigma_r = f' \,\,\Lambda \,;\quad \qquad \Sigma_\chi = - 2\,\sin \frac{f}{2}\,P^{(+)}_\chi\,;\quad\qquad \Sigma_{\bar{\chi}} =  2\,\sin \frac{f}{2}\,P^{(-)}_{\bar{\chi}} \lab{sigmasrx}
\er
On the other hand, using \rf{pir} the topological charge \rf{charge3d} becomes
\br
 Q&=& \frac{1}{2\,\pi}\,\left[f\(r\)-\sin f\(r\)\right]_{r=0}^{r=\infty}\;Q_{{\rm top}} \;\qquad\quad {\rm where}\quad \nonumber\\
 Q_{{\rm top}} & = & \eta\,\frac{i}{4\,\pi\,\kappa}\, \int dz\,d\bar{z}\;{\rm Tr}\(P^{(+)}_\chi\,P^{(-)}_{{\bar \chi}}\) \qquad\quad {\rm with}\quad\quad \eta \equiv \left\{\begin{array}{ll} +1, & \chi = z\\ -1, & \chi=\bar{z}\end{array} \right.
 \lab{pretopcharge3db}
\er

Finally, using \rf{gdef2} we have that the Lie algebra element appearing in the ansatz \rf{holog} is given by 
\be
g\,\Lambda\,g^{-1}=\Lambda -\frac{1}{\(1+\omega\)}\left(\sbr{S}{S^{\dagger}}+i\(S-S^{\dagger}\)\right)
\lab{bitofanasatz}
\ee
In particular, if there is a real number $c$ such that 
\be 
\Lambda^2=c\,\Lambda + \frac{1}{4}\,\(1-c^2\)\, \one 
\lab{lambdacond}
\ee
it  follows by construction that
\be Z \equiv \frac{1+c}{2}\,\one -g\,\Lambda\,g^{-1} = \frac{1+c}{2}\,\one -\Lambda +\frac{1}{\(1+\omega\)}\left(\sbr{S}{S^{\dagger}}+i\(S-S^{\dagger}\)\right) \lab{zdef}\ee
is a projector, i.e. $Z^2=Z$. Therefore, for such a case the rational map \rf{holog}  becomes
\be 
U= e^{if\,\frac{1+c}{2}\,\one}\,e^{-i\,f\,Z}=e^{if\,\frac{c+1}{2}}\,\left[\one+\(e^{-if}-1\)Z\right]
\lab{Uexpanded}\ee

\section{Examples}
\label{sec:examples}
\setcounter{equation}{0}

We now give explicit constructions of the rational ansatz \rf{holog} for the some hermitian symmetric spaces listed in Table \ref{listhermitian}. 

\subsection{The case of $CP^1=SU(2)/U(1)$}

In the case of $CP^1=SU(2)/U(1)$ we use the spinor representation of $SU(2)$ and take the matrix $S$ as
\br
S=\(\begin{array}{cc}
0& u\\
0&0
\end{array}\) \lab{ssu2}
\er
with $u$ being a complex field parameterizing $CP^1$. From \rf{finalholos} and \rf{finalantiholos} we see that $u$ must be holomorphic, i.e. $u=u\(z\)$, or anti-holomorphic, i.e. $u=u\({\bar z}\)$ (see \rf{coordinates}). 
Such a matrix satisfies the conditions \rf{omegadef} with 
\be
\omega=  u\, {\bar u}\equiv \mid u\mid^2
\ee
The group element \rf{gdef2} becomes
\br
g=\frac{1}{\sqrt{1+\mid u\mid^2}}\,\(
\begin{array}{cc}
1&i\, u\\
i\,{\bar u}& 1
\end{array}\)
\er
and \rf{bitofanasatz} becomes
\br
g\,\Lambda\,g^{-1}=\frac{1}{2}\,\frac{1}{\(1+\mid u\mid^2\)}\,\(
\begin{array}{cc}
1-\mid u\mid^2&-2\,i\, u\\
2\,i\,{\bar u}& -1+\mid u\mid^2
\end{array}\)\equiv \frac{1}{2}\,{\hat n}_a\,\sigma_a\,;\qquad\quad
\Lambda=\frac{1}{2}\,\sigma_3
\er
where $\sigma_a$, $a=1,2,3$, are the Pauli matrices, and ${\hat n}$ is the unit vector
\be
{\hat n}=\frac{1}{\(1+\mid u\mid^2\)}\,\(2\,u_2\,,\, 2\,u_1\,,\, 1-\mid u\mid^2\)\,; \qquad\quad u=u_1+i\,u_2\,;\qquad\quad {\hat n}^2=1
\ee
Therefore, the ansatz \rf{holog} becomes 
\be
U= e^{i\,f\(r\)\, {\hat n}\cdot{\vec\sigma}/2}
\lab{hologsu2}
\ee
which is the usual rational map ansatz used in the literature \cite{mantonbook,rational1,rational2}, where $u$ is taken as the ratio of two polynomials on the variable $z$ (or ${\bar z}$) \cite{mantonbook}. 

\subsection{The case of $CP^N=SU\(N+1\)/SU\(N\)\otimes  U\(1\)$}

In the case of $CP^N=SU\(N+1\)/SU\(N\)\otimes U\(1\)$, corresponding to the Type I case of Table \ref{listhermitian} with $p=N$ and $q=1$, we choose $\alpha_*=\alpha_N$, which implies $\lambda_*=\lambda_N$, the
fundamental $(N+1)\times (N+1)$ representation of $SU(N+1)$ and take the matrix $S$ as 
\be
\lab{S3a}
 S=\(\begin{array}{cc} O_{N\times N} & u \\ O_{1\times N} & 0 \end{array} \) 
  \ee
with the $CP^N$ being parameterized by $N$ complex scalar fields $u_a$, where $a=1,\,...N$, corresponding with the components of $u^T=\(u_1,\,...,\,u_N\)$. In addition, $O_{1\times N}$ is a $1\times N$ zero matrix, and so on. From \rf{finalholos} and \rf{finalantiholos} we see that $u_a$ must be holomorphic, i.e. $u_a=u_a\(z\)$, or anti-holomorphic, i.e. $u_a=u_a\({\bar z}\)$  (see \rf{coordinates}). Such a matrix satisfies the conditions \rf{omegadef} with 
\be
\omega=  u^\dagger \, u =  \mid u\mid^2 
\ee
The $S$ matrix \rf{S3a} also parametrizes  ${\cal P}_+$ with the two sets of $N$ generators of the abelian subspaces ${\cal P}_+$ and ${\cal P}_- $  being given respectively by 
\be \(E_{+\alpha_\kappa}\)_{bc} \equiv \delta_{b\,\kappa}\,\delta_{c\,(N+1)} \, \quad \qquad {\rm and}\quad \qquad \(E_{-\alpha_\kappa} \)_{bc}= \(E_{+\alpha_\kappa}^{\dagger}\)_{bc} = \delta_{b\,(N+1)}\,\delta_{c\,\kappa} \lab{p+-}\, \:\:\: \ee 
where $b,\,c = 1,...,\, N+1,$ and $\kappa =1,...,\, N$.

The group element \rf{gdef2} and the matrix $\Lambda$ defined in \rf{sigmadef} corresponds respectively to \cite{erica} 
\be g=\frac{1}{\vartheta}\(\begin{array}{cc}
\Delta & iu \\iu^\dagger & 1
\end{array}\) \,;\qquad\quad \quad \qquad \Lambda=\frac{1}{N+1}\(\begin{array}{cc}
\one_{N \times N} & 0 \\0 & -N
\end{array}\)\lab{mat22}
\ee
where $\Delta$ is a $N\times N$ Hermitian invertible matrix defined by 
\be \Delta_{ij}\equiv \vartheta \,\delta_{ij}-\frac{u_i \,u_j^*}{1+\vartheta}\,; \qquad\Delta_{ij}^{-1}=\frac{1}{\vartheta} \,\(\delta_{ij}+\frac{u_i \,u_j^*}{1+\vartheta}\)\, \qquad {\rm with } \qquad \vartheta \equiv \sqrt{1+u^\dagger u} \lab{deltadef}\ee
where $i,\,j =1,\,...,\,N$. Note that due to \rf{deltadef} $u$ is an eigenvector of $\Delta$ with eigenvalue $+1$, i.e. $\Delta u = u$, which also implies that $u^\dagger\Delta =u^\dagger$. See \cite{erica} for details. The $\Lambda$ matrix given in \rf{mat22} satisfies the condition \rf{lambdacond} for $c=\frac{1-N}{1+N}$, then  \rf{bitofanasatz} becomes
\be 
g\,\Lambda \,g^{-1}=  \frac{\one}{N+1}-Z\qquad {\rm with} \qquad Z \equiv \frac{V \otimes {\bar V}}{\mid V \mid^2}= \frac{1}{u^\dagger u +1}\,\(\begin{array}{cc}
u\otimes \bar{u} & i\,u \\-i\,u^\dagger & 1 \end{array}\) \lab{projector}
\ee
where $V\equiv \(i\,u,\,1\)^T$ and $Z$ introduced in \rf{zdef} is a $(N+1)\times (N+1)$ hermitian projector that acts as a map $S^2 \rightarrow CP^{N}$. From \rf{projector} we observe that our generalized rational map ansatz \rf{holog}, for the $CP^N$ case, is equivalent to the construction of Ioannidou et al. ansatz proposed in \cite{ioannidou1,ioannidou2}, since using \rf{Uexpanded} the field $U$ takes the form
\be U=e^{i\,f(r)\,\({\scriptsize \one}/(N+1)-Z\)}=e^{i\,f(r)/(N+1)}\,\left[\one +\(e^{-i\,f(r)}-1\)\,Z\right] \lab{Wansatz}\ee

\subsection{The case of $SU\(p+q\)/SU\(p\)\otimes SU\(q\)\otimes U\(1\)$}
\label{Spq}

In the case of $SU\(p+q\)/SU\(p\)\otimes SU\(q\)\otimes U\(1\)$, corresponding to the Type I case of Table \ref{listhermitian}, we choose $\alpha_*=\alpha_p$, which implies $\lambda_*=\lambda_p$, the fundamental $(p+q)\times (p+q)$ representation of $SU(p+q)$ and take the matrix $S$ as 
\br
& S=\(\begin{array}{cc} O_{p\times p} & u \otimes v\\ O_{q\times p} & O_{q\times q}\end{array} \) \lab{Spq}
\er
with $SU\(p+q\)/SU\(p\)\otimes SU\(q\)\otimes U\(1\)$ being parameterized by $p$ complex scalar fields $u_a$ and $q$ complex scalar fields $v_b$, where $a=1,\,...,\,p$ and $b=1,\,...,\,q$, corresponding with the components of $u^T=\(u_1,\,...,\,u_p\)$ and $v^T=\(v_1,\,...,\,v_q\)$. In addition, $O_{1\times N}$ is a $1\times N$ zero matrix, and so on. From \rf{finalholos} and \rf{finalantiholos} we see that the fields $u_a$ and $v_a$ must be holomorphic, i.e. $u_a=u_a\(z\)$, and $v_a=v_a\(z\)$, or anti-holomorphic, i.e. $u_a=u_a\({\bar z}\)$, and $v_a=v_a\({\bar z}\)$ (see \rf{coordinates}). Such a matrix satisfies the conditions \rf{omegadef} with 
\be
\omega=   \(u^\dagger\,u\)\,\(v^\dagger\,v\)=  \mid u\mid^2 \,\mid v\mid^2
\ee
and the value of the $\kappa$ factor introduced in \rf{normalizedtr} is 
\be
\kappa = \frac{1}{2} \lab{kappa}
\ee
The group element \rf{gdef2} becomes
\be
g=\frac{1}{\vartheta}\,\(\begin{array}{cc} \Delta_u & i\,u \otimes v\\ i\,\overline{v}\otimes {\bar u} & \Delta_v^T
\end{array} \lab{gpq}
\)
\ee
with $\vartheta = \sqrt{1 + (u^\dagger u)\,(v^\dagger v) }$, $\Delta_x \equiv \vartheta \one + \(1-\vartheta\)\, T_x$, where $T_x\equiv \frac{x\otimes {\bar x}}{x^\dagger \cdot x}$ is a projector and $x$ stands for $u$ and $v$.  Note that by construction, $u$ and $\overline{v}$ are eigenvectors respectively of $\Delta_u$ and $\Delta_v^T$ with eigenvalues $+1$, i.e. $\Delta u = u$ and $\Delta_v^T\,\bar{v}= \bar{v}$. On the other hand, the inverse of $\Delta_x$ is given by $\Delta_x^{-1} = \vartheta^{-1}\,\( \one - \(1-\vartheta\)\, T_x\) $, and we also have $v^T\,\Delta_v^T=v^T$ and  $u^\dagger\,\Delta_u=u^\dagger$. 
The matrix $\Lambda$ and the $e^{\pm i \,\pi\,\Lambda}$ factor that characterizes the involutive automorphism $\sigma$, both defined in \rf{sigmadef}, corresponds respectively to\footnote{The inverse of the Cartan matrix of the $SU(p+q)$ Lie group can be written as $K^{-1}_{ij}=\min\(i,\,j\)  - i\,j/(p+q)$.  All the roots have the same size and we use the normalization $\alpha_1^2=2$, and so the matrices $H_{\alpha_a}\equiv\frac{2\,\alpha_a\cdot H}{\alpha_a^2}$ become $\(H_{{\alpha_b}}\)_{ij} =\delta_{ib}\,\delta_{jb}-\delta_{i(b+1)}\,\delta_{j(b+1)}$. Thus, the choice $\alpha_*=\alpha_p$, which implies $\lambda_*=\lambda_p$, fixes $\Lambda= K_{*b}^{-1}\,H_{\alpha_b}$, i.e.  $\Lambda_{ij}=\left[\min\(p,\,i\)- \min\(p,\,i-1\)-\frac{p}{p+q}\right]\,\delta_{ij}$. Clearly, the  non-vanishing entries of $\Lambda$ are $\Lambda_{ii}=\frac{q}{p+q}$ for $i\leq p$, and  $\Lambda_{ii}=-\frac{p}{p+q}$ for $i\geq p +1$.}
\be
\Lambda = \frac{1}{p+q}\,\(\begin{array}{cc} q\,\one_{p\times p} & O_{p\times q}\\ O_{q\times p} & -p\,\one_{q\times q}\end{array}\)\,;  \quad\qquad e^{i\,\pi\,\Lambda}=e^{-\frac{i\,\pi\,p}{p+q}}\(\begin{array}{cc} -\one_{p\times p} & O_{p\times q}\\ O_{q\times p} & \,\one_{q\times q}\end{array}\) \lab{lambdapq}
\ee

The $\Lambda$ matrix given in \rf{lambdapq} satisfies the condition \rf{lambdacond} for $c=\frac{q-p}{q+p}$, then  \rf{bitofanasatz} becomes 
\be 
g\,\Lambda \,g^{-1}=  \Lambda  - \frac{1}{\vartheta^2}\,\(\begin{array}{cc} \omega\, T_u & i\,u\otimes v \\-i\,\bar{v}\otimes {\bar u} & -\omega\, T_v^T \end{array}\) = \frac{q}{p+q}\,\one-Z \lab{glambdapq}
\ee
where $Z$ introduced in \rf{zdef} is the  projector
\be Z \equiv \frac{1}{\vartheta^2}\,\(\begin{array}{cc} \omega\, T_u & i\,u\otimes v \\-i\,\bar{v}\otimes {\bar u} & \(\Delta_v^{T}\)^2 \end{array}\)\ee
Using \rf{glambdapq} or \rf{Uexpanded}, our generalized rational map ansatz \rf{holog} for the field $U$ takes the form
\be U=e^{\frac{i\,q\,f(r)}{p+q}}\,\left[\one +\(e^{-i\,f(r)}-1\)\,Z\right] \lab{upq}\ee 
Clearly, the equation \rf{glambdapq} is reduced to \rf{projector} for the $CP^N$ case, i.e. for $p=N$ and $q=1$ with $v=v_1$ implying $\Delta_v=1$. However, in such a case the field $v_1$ can be absorbed in the field $u$ through the transformation $u\rightarrow u/v_1$, which transforms $\Delta_u$ to \rf{deltadef} and $g$ to \rf{mat22}. Therefore, such a case is equivalent to impose that the field $v_1$ is just the unit number, i.e. $v=1$.

\subsection{The case of $Sp\(N\)/SU\(N\)\otimes U\(1\)$}

In the case of $Sp\(N\)/SU\(N\)\otimes U\(1\)$ corresponding to the Type IV case of Table \ref{listhermitian}, we can choose  $\alpha_*=\alpha_N$, which implies $\lambda_*=\lambda_N$. The $N$ simple roots of $Sp(N)$ can be written as   
\br
\alpha_1 &=&\frac{1}{\sqrt{2}}\(1,\,-1,\,...,\,0\),\qquad \alpha_{N-1} =\frac{1}{\sqrt{2}}\,\(0,\,...,\,1,\,-1\),\qquad \alpha_N = \(0,\,0,\,...,\,\sqrt{2}\) \lab{spnroots}\qquad 
\er
Clearly, $\alpha^2_1=\,...=\alpha^2_{N-1}=1$, $\alpha^2_{*}=\alpha^2_{N}=2$. We choose the $(2N)\times (2N)$ representation of $Sp(N)$, with the generators of the Cartan subalgebra being given by
\be \(H_i\)_{ab} = \frac{1}{\sqrt{2}}\,\delta_{ab}\,\(\delta_{ai}-\delta_{a(i+N)}\),\qquad\qquad i=1,\,...,\,N \ee
while the $N\,(N+1)/2$ generators of the ${\cal P}_+$ subalgebra corresponds to 
\be
\lab{Pabs} P_{ab}^{(+)} \equiv \( \begin{array}{cc} O_{N\times N} & B_{ab}\\O_{N\times N} & O_{N\times N}\end{array}   \); \quad {\rm with} \quad \(B_{ab}\)_{ij}=\left\{ \begin{array}{ll} \delta_{ai}\,\delta_{bj} ,& {\rm if }\,a=b\\ \delta_{ai}\,\delta_{bj} +\delta_{aj}\,\delta_{bi} , &{\rm otherwise }\end{array}   \right.
\ee
where $a=1,\,...,\,b$ and  $b=1,\,...,\,N$. The fundamental weights and simple roots are related by $\lambda_a=K^{-1}_{ab}\,\alpha_b$, with $K_{ab}$ being the Cartan matrix. The $N$-th row of the inverse of the Cartan matrix is given by $ \(1,\,2,\,...,\,N-1,\,  N/2\)$, and so from  \rf{spnroots} we get $\lambda_*  = \frac{1}{\sqrt{2}}\,\(1,\,...,\,1,\,1\)$, fixing $\Lambda$ given in \rf{sigmadef} through
\br
\Lambda &=& \frac{H_1 +...+H_N}{\sqrt{2}}=\frac{1}{2}\,\(\begin{array}{cc} \one_{N\times N} & O_{N\times N}\\ O_{N\times N} & -\one_{N\times N}\end{array}\) \lab{lambdaspn}
\er

The generator $\Lambda$ given in \rf{lambdaspn} for $G=Sp(N)$ and in \rf{lambdapq} with $p=q=N$ for $G=SU(2\,N)$, leads to the same involutive automorphism $\sigma$, as defined in \rf{sigmadef}.\footnote{Note that any of the generators of the ${\cal P}_+$ subalgebra of $Sp(N)$, given in \rf{Pabs}, can be written as a linear combination of the $p\cdot q $ generators $E_{ij}$, where $i=1,\,...,\,p$ and $j=p+1,\,...,\,p+q$, of the ${\cal P}_+$ subalgebra of $SU(2\,N)$ used in the section \ref{Spq}, which in turn can be written as $\(E_{ij}\)_{ab} \equiv \delta_{a\,i}\,\delta_{b\,j}$, with $a,\,b = 1,...,\, p+q$.} 
Our ansatz must be of the form $S\equiv w_{ab}\,P_{ab}^{(+)}$, where $w_{ab}$, with $a=1,..., \,b$ and $b=1,...,N$, forms a set of $N\,(N+1)/2$ complex scalar fields. This satisfies \rf{omegadef2} if we impose that $S$ can also be written as \rf{Spq} for the case of $SU(2\,N)/SU(N)\otimes SU(N)\otimes U(1)$, which is equivalent to impose that $u\otimes v$ must be symmetric. This condition can be satisfied by $v=v_1\,\frac{u}{u_1}$, which fix the $N-1$ fields $v_2,\,...,\,v_N$ in terms of the fields $v_1$ and $u$, implying that $T_v=T_u$ and $\Delta_v=\Delta_u$. Thus, the matrix $S$ for the case of ${Sp\(N\)/SU\(N\)\otimes U\(1\)}$ takes the form 
\be 
S=\frac{v_1}{u_1}\,\(\begin{array}{cc} O_{N\times N} & u \otimes u\\ O_{N\times N} & O_{N\times N}\end{array} \)   \qquad {\rm with} \qquad   \omega=\left| \frac{v_1}{u_1}\right|^2  \,\(u^\dagger\,u\)^2 \lab{S3d}  \quad  
\ee
The  group element \rf{gdef2} becomes 
\be
g =   \frac{1}{\vartheta}\, \(\begin{array}{cc} \Delta_u & i\,\frac{v_1}{u_1}\,u \otimes u \\ i\,\frac{\bar{v}_1}{\bar{u}_1}\,\bar{u} \otimes {\bar u}  & \Delta_{u}^T \end{array} \) 
\lab{gspna}
\ee
with $\vartheta = \sqrt{1+\omega}$ and \rf{bitofanasatz} is given by
\be 
g\,\Lambda \,g^{-1}= \Lambda  - \frac{1}{\vartheta^2}\,\(\begin{array}{cc} \omega\, T_u & i\,\frac{v_1}{u_1}\,u \otimes u \\ -i\,\frac{\bar{v}_1}{\bar{u}_1}\,\bar{u} \otimes {\bar u} & -\omega\, T_u^T \end{array}\) = \frac{1}{2}\,\one-Z \lab{glambdaspn}
\ee
where $Z$ is the projector
\be Z \equiv \frac{1}{\vartheta^2}\,\(\begin{array}{cc} \omega\, T_u & i\,\frac{v_1}{u_1}\,u \otimes u \\- i\,\frac{\bar{v}_1}{\bar{u}_1}\,\bar{u} \otimes {\bar u} & \(\Delta_u^{T}\)^2 \end{array}\)\ee
Using \rf{glambdapq}, our generalized rational map ansatz \rf{holog} for the field $U$ takes the form
\be U=e^{\frac{i\,f(r)}{2}}\,\left[\one +\(e^{-i\,f(r)}-1\)\,Z\right] \lab{uspn}\ee 

\section{Approximation for the $SU(N)$ Skyrmions}
\label{sec:applications}
\setcounter{equation}{0}

The standard Skyrme model is a nonlinear $SU(N)$ sigma model field theory defined in $(3+1)$-dimensional Minkowski space by the action \cite{ioannidou1, ioannidou2}
\be S_{{\rm Skyrme}}=\int d^4x\left[\frac{f_\pi^2}{16}\Tr \left(R_\mu R^\mu\right)-\frac{1}{32\,e_0^2}\Tr \left([R_\mu ,R_\nu][R^\mu ,R^\nu]\right)\right]\lab{actiona} \ee
where $R_\mu=i\,\pa_\mu U\,U^{-1}$, the so-called Skyrme field $U$ maps the three-dimensional physical space to $SU(N)$, and the metric is given by $ {\rm diag.}\(1,\,-1,\,-1,\,-1\)$. In natural units, the pion decay constant $f_\pi$ has the dimension of mass, while $e$ is a dimensionless coupling constant.  We choose the scale and energy units, respectively, as $2/(e_0 \, f_\pi)$ and $12\,\pi^2\,f_\pi/(4 \, e_0)$, where we have introduced the usual factor $12\,\pi^2$ in the energy scale. Using also the metric \rf{metric}, the static energy associated with the action \rf{actiona} becomes 
\br
E & = & \frac{1}{12\,\pi^2}\,\int d^3x \({\cal E}_2+ {\cal E}_4\) \;;\nonumber \qquad\qquad{\rm with} \qquad\\
{\cal E}_2 & = &  \frac{1}{2}\,g^{\alpha\beta}\,{\rm Tr} \(\Sigma_\alpha \, \Sigma_\beta\) =\frac{1}{2}\, {\rm Tr}  \( \Sigma_r \, \Sigma_r\) + g^{\chi\bar{\chi}}\, {\rm Tr}  \( \Sigma_\chi \, \Sigma_{\bar{\chi}}\) \nonumber\\
{\cal E}_4 & = &  -\frac{g^{\alpha\beta}\,g^{\gamma\delta}}{16}\,{\rm Tr} \(\left[\Sigma_\alpha ,\, \Sigma_\gamma\right]\,\left[\Sigma_\beta ,\, \Sigma_\delta\right]\) =  -\frac{ g^{\chi\bar{\chi}}}{8}\,{\rm Tr}\( 2\,\left[\Sigma_r ,\, \Sigma_\chi\right]\,\left[\Sigma_r ,\, \Sigma_{\bar{\chi}}\right]-g^{\chi\bar{\chi}}\,\left[\Sigma_\chi ,\, \Sigma_{\bar{\chi}}\right]^2\) \nonumber \\
&=&  \frac{ g^{\chi\bar{\chi}}}{8}\,{\rm Tr}\( 2\,f^{\prime 2}\, \Sigma_\chi\, \Sigma_{\bar{\chi}}+g^{\chi\bar{\chi}}\,\left[\Sigma_\chi ,\, \Sigma_{\bar{\chi}}\right]^2\) \nonumber
\er
where $g^{\chi\bar{\chi}} = \frac{\(1+\mid z\mid^2\)^2}{2\,r^2}$ and we use $[\Sigma_r,\, \Sigma_\chi] = f'\,\Sigma_\chi$ and $[\Lambda,\, \Sigma_{\bar{\chi}}] = -f'\,\Sigma_{\bar{\chi}}$, which is a consequence of \rf{piplusminusdef} and \rf{sigmasrx}. Using the $S$ and $\Lambda$ matrices for the coset $SU(p+q)/SU(p)\otimes SU(q)\otimes U(1)$, as given in \rf{Spq} and \rf{lambdapq}, together with \rf{sigmasrx}, we obtain 
\be {E} = \frac{1}{12\,\pi^2}\,\int d^3x \,\left[\frac{f^{\prime 2}}{4}\, \alpha_{pq} +2\,\frac{\sin^2 \frac{f}{2}}{r^2}\,\(1+\frac{f^{\prime 2}}{4}\)\,F_2 + \frac{\sin^4 \frac{f}{2}}{r^4}\,F_4 \right]\lab{energyminpre}
\ee
with 
\be
F_2\equiv -\tilde{g}\,{\rm Tr}\( P_\chi^{(+)} \, P_{\bar{\chi}}^{(-)}\) \;; \qquad F_4\equiv \frac{\tilde{g}^2}{2}\,\,{\rm Tr}\(\left[P_\chi^{(+)} ,\, P_{\bar{\chi}}^{(-)}\right]^2\) \;; \qquad \alpha_{pq}\equiv   \frac{2\,p\,q}{p+q}   \lab{conditionsp}
\ee
where $\tilde{g}\equiv \(1+\mid z\mid^2\)^2$ and $P_\chi^{(+)}$ and $P_{\bar{\chi}}^{(-)}$ are given by \rf{pir} and \rf{Spq}. In the case where the real-valued angular functions  $F_2$ and $F_4$ are both constants, the static energy density associated with \rf{energyminpre} became spherically symmetric. These two conditions can be both satisfied for the following choice of the $u$ and $v$ fields in \rf{Spq} given by
\br
u_c&=& u_1=\frac{p_u\(\chi\)}{q_u\(\chi\)};\qquad  v_d= v_1=\frac{p_v\(\chi\)}{q_v\(\chi\)}; \qquad  w\equiv \sqrt{p\,q}\,u_1\,v_1 \lab{restrictive}\er 
for all $c=1,\,...,\,p$ and $d=1,\,...,\,q$,  together with 
\br
w=e^{i\,\alpha}\,\chi \qquad\qquad {\rm or} \qquad\qquad  w=\frac{\beta\(\chi-\mid \beta \mid^{-1}\,e^{i\alpha}\)}{\chi+\mid \beta \mid\,e^{i\alpha}} \lab{restrictive2}
\er
where $\alpha$ is a real constant contained in the interval $\left[0,\,2\pi\)$, and $\beta$ is an arbitrary complex constant with $\beta \neq 0$. In addition, $p_u$ and $q_u$ are two polynomials with no common roots, and the same holds for $p_v$ and $q_v$. Therefore, we can write the field $w$ as a rational map, meaning it can also be written as the ratio of two polynomials, $p_w(\chi)$ and $q_w(\chi)$ without common roots. However, we must restrict to the cases where $p_u$ and $q_v$  do not share common roots, and the same holds for $p_v$ and $q_u$.

Note that the choice \rf{restrictive2} corresponds to the cases where the rational map ansatz reduces the Euler-Lagrange equations of the standard $SU(2)$ Skyrme model to a single radial equation \cite{mantonbook}. In addition, the map $w$ between two spheres, given in \rf{restrictive2}, have topological degree ${\rm deg}\, w=1$, and using \rf{normalizedtr} and \rf{kappa} the topological charge \rf{pretopcharge3db} becomes
\br
 Q&=& \left[\frac{f\(r\)-\sin f\(r\)}{2\,\pi}\right]_{r=0}^{r=\infty} Q_{{\rm top}} \,;\qquad\quad  Q_{{\rm top}} \equiv \eta \, {\rm deg}\,w= \eta \lab{charge1pre}
\er

Let us introduce $I_{p\times q}$ as a $p\times q$ matrix with all entries equal to $1$, which satisfies $I_{p\times q}\,\(I_{p\times q}\)^\dagger = q\,I_{p\times p}$ and $\(I_{p\times q}\)^\dagger\,I_{p\times q} = p\,I_{q\times q}$. Using \rf{pir}, \rf{Spq} and  the ansatz \rf{restrictive}, but not the explicit form of the field $w$ given in \rf{restrictive2}, we obtain 
\br 
S & =& \frac{w}{\sqrt{p\,q}} \, \(\begin{array}{cc} O_{p\times p} & I_{p\times q}\\ O_{q\times p} & O_{q\times q}\end{array}\)\, \qquad \Rightarrow \qquad P_\chi^{(+)} =  i\,\frac{\vartheta^{-2}}{\sqrt{p\,q}} \, \pa_\chi w \,\(\begin{array}{cc} O_{p\times p} & I_{p\times q}\\ O_{q\times p} & O_{q\times q}\end{array}\) 
\er
leanding to 
\be
F_2= \(\frac{1+\mid \chi \mid^2}{1+\mid w \mid^2}\)^2 \,\pa_\chi w\,\pa_{\bar{\chi}} \bar{w} \; \qquad\quad  {\rm and } \quad\qquad F_4= F_2^2   \lab{conditionspp}
\ee
As shown in \cite{quasiself},  the rational maps \rf{restrictive2} are the only rational maps that satisfy the condition $\(\frac{1+\mid \chi \mid^2}{1+\mid w \mid^2}\)^2 \,\pa_\chi w\,\pa_{\bar{\chi}} \bar{w} =const.$, and for such maps, we have $F_2=F_4=1$.  Therefore, the choice \rf{restrictive}-\rf{restrictive2} leads to a spherically symmetric static energy density, and \rf{energyminpre} becomes
\br  E_{pq} &=&  \alpha_{pq}\,E_a+E_b \;;\qquad\qquad{\rm with} \qquad \lab{energymin}\\
 E_a &\equiv & \frac{1}{3\,\pi}\,\int dr\,r^2\,\frac{f^{\prime 2}}{4}  \; \qquad\quad  E_b \equiv \frac{1}{3\,\pi}\,\int dr\,r^2\,\left[2\,\frac{\sin^2 \frac{f}{2}}{r^2}\,\(1+\frac{f^{\prime 2}}{4}\) + \frac{\sin^4 \frac{f}{2}}{r^4} \right] \nonumber  
\er

Although the static energy \rf{energymin} is symmetric under the exchange $p \leftrightarrow q$, distinct values of $p$ and $q$ can lead to distinct energy functionals \rf{energymin} for the same value of $N = p + q$.  Therefore, we must determine the values of $p$ and $q$ satisfying $p + q = N$ that provide the best (lowest-energy) approximation for the $SU(N)$ Skyrmions. Fixing $q = N - p$, the factor $\alpha_{pq}$ is reduced to the quadratic polynomial in $p$, given by $\alpha_{pq}= \frac{2}{N}\,p\,\(N-p\)$. Thus, since $p$ lies in the interval $1, \,...,\, N - 1$, the global minimum of $\alpha_{pq}$ for each value of $N$, denoted by $\alpha_N^{{\rm min}}$, occurs for $p = N - 1$ and $q = 1$, or vice versa. This corresponds to the $CP^{N-1} = SU(N)/SU(N-1) \otimes U(1)$ case and leads to $\alpha_N^{{\rm min}}=\alpha_{(N-1)\,1}=2\,\(1-\frac{1}{N}\)$.

The static energy associated with the minima of \rf{energymin} increases as $\alpha_{pq}$ increases for the non-trivial topological solutions $Q\neq 0$. Therefore, our ansatz provides a better approximation of the $SU(N)$ Skyrmions when the principal variable parametrizes the coset $CP^{N-1}$, where the values of $\alpha_{pq}$, restricted by $p + q = N$, attain their minimum $\alpha_N^{{\rm min}}$. The corresponding static energy \rf{energymin} is then reduced to $E_N^{{\rm min}}\equiv \alpha_N^{{\rm min}}\,E_a+E_b $.

The proof of the statement above is straightforward. First, since $Q \neq 0$, the profile function cannot be constant over the entire physical space, then $\int r^2 f^{\prime 2 }>0$ which  implies that $E_a > 0$. Let us denote $f_{pq}$ and $f_{lk}$ as the global energy minimizers of $E_{pq}$ and $E_{lk}$, respectively, for any positive integer values of $p, q, l,$ and $k$. We want to prove that if $\alpha_{pq} > \alpha_{lk}$, then it follows that $E_{pq}\left[f_{pq},\,f_{pq}'\right] > E_{lk}\left[f_{lk},\,f_{lk}'\right] $. Assuming that $\alpha_{pq} > \alpha_{lk}$, it follows that there exists a positive real number $\epsilon$ such that $\alpha_{pq} = \alpha_{lk} + \epsilon$. Taking an arbitrary  fixed profile function $f=f_F$ and using \rf{energymin} and $E_a \left[f_F'\right]>0$, we have
\be E_{pq}\left[f_F,\,f_F'\right] = E_{lk}\left[f_F,\,f_F'\right] + \epsilon\,E_a\left[f_F'\right] > E_{lk}\left[f_F,\,f_F'\right] \geq E_{lk}\left[f_{lk},\,f_{lk}'\right] \lab{ppp1} \ee
Finally, by choosing the arbitrary profile function $f_F = f_{pq}$ in \rf{ppp1} we obtain  $E_{pq}\left[f_{pq},\,f_{pq}'\right] > E_{lk}\left[f_{lk},\,f_{lk}'\right] $, completing the proof.

The static energy functional \rf{energymin} corresponding  to the $CP^{N-1}$ case can be written as
\br \quad E_N^{{\rm min}}& =& E_N^2 + E_N^4\qquad \quad{\rm with} \quad\qquad \lab{energymin2} E_N^2 \equiv \frac{1}{3\,\pi}\,\int dr\,r^2\,\left[ \frac{f^{\prime 2}}{4}\, \alpha_N^{{\rm min}} +2\,\frac{\sin^2 \frac{f}{2}}{r^2}\right] \qquad \lab{enermin2}\\
 E_N^4 &\equiv &\frac{1}{3\,\pi}\,\int dr\,r^2\,\left[\frac{\sin^2 \frac{f}{2}}{r^2}\,\frac{f^{\prime 2}}{2} + \frac{\sin^4 \frac{f}{2}}{r^4} \right]\; \nonumber
\er
which leads to the following Euler-Lagrange equation
\be
\Delta f  \equiv \frac{1}{2}\,\(\alpha_N^{{\rm min}}\,r^2+ 2\,\sin^2 \(\frac{f}{2}\)\)\,f''  +\alpha_N^{{\rm min}}\,r\,f'  +\sin f\,\(\frac{f^{\prime 2}}{4} -1 -\frac{\sin^2 \frac{f}{2}}{r^2} \)= 0 \lab{eqf}
\ee

Since $\alpha_N^{{\rm min}}$ is monotonically increasing in $N$, the energy $E_N^{{\rm min}}$ of the $SU(N)$ Skyrmions (minimizers of \rf{energymin}) is also monotonically increasing with $N$. Thus, there is a finite upper bound for such energies, given by $E_\infty^{{\rm min}}$, which corresponds to the energy of the minimizer of \rf{energymin2} with $\alpha_\infty^{{\rm min}} \equiv \lim_{N \rightarrow \infty} \alpha_N^{{\rm min}} = 2$. Fixing the boundary conditions $f(0)=2\,\pi$ and $f(\infty)=0$ the topological charge \rf{charge1pre} become $Q=\eta$.

The interesting relation $\alpha_{22}=\alpha_N^{{\rm min}}=2$ implies that the energy minimizers of \rf{energymin} for $p=q=2$ $(N=4)$ have the same energy of the minimizers of \rf{energymin2} for $N\rightarrow \infty$. This give us the ideia of how much the energy $E_{pq}$ of the minimizers of \rf{energymin} can grows in relation to the minima of $E_N^{{\rm min}}$ when we varry the values of $p$ and $q$ while keeping the value of its sum $N$ fixed. Fixing the boundary conditions $f(0)=2\,\pi$ and $f(\infty)=0$ the topological charge \rf{charge1pre} become $Q=\eta$.

The profile function associated with solutions of \rf{eqf} falls asymptotically slowly with $r^{-1}$, being convenient to compactify the radial interval $\IR_+$ using the variable $x\equiv \frac{r}{2+r} \in [0,\,1]$. We obtain the energy minimizers of \rf{energymin2} by solving the compactified version of \rf{eqf} with these boundary conditions for $N=2,\,...,\, 100$ and for $N\rightarrow \infty$ ($E_\infty^{{\rm min}}$) using the gradient flow method on the compactified grid with spacing $1/600$ and $n_p=601$ points. The derivatives are computed with fourth-order central formulas, and the maximum value of $\mid \Delta f \mid$ for the solutions is lower than $10^{-4}$. On the other hand, due to Derrick's scaling argument, we must have $E_N^2 = E_N^4$, and numerically, $D \equiv \mid E_N^2 - E_N^4\mid /E_N^{{\rm min}}$ is lower than $2 \times 10^{-5}$ for all the solutions.
\begin{figure}[htp!]
\begin{center}
\includegraphics[scale=0.42]{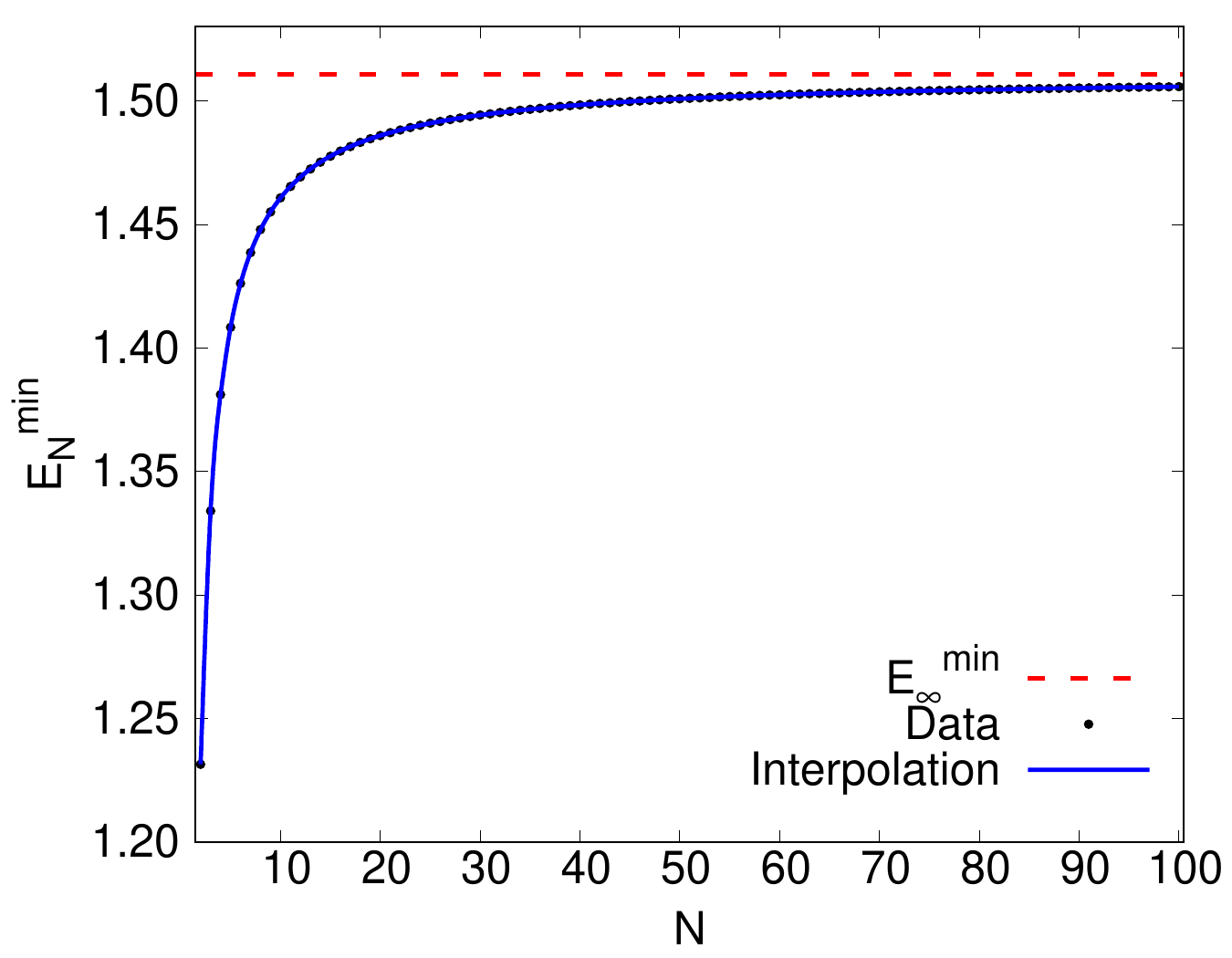}
\end{center}
\caption{The energy $E_N^{\rm min}$ of the minimizers of \rf{energymin2} corresponds to spherically symmetric Skyrmions with $\mid Q\mid=1$. The black dots correspond to the numerical data for $N=2,\,...,\,100$, the solid blue line represents the data interpolation, and the dashed red line corresponds to the upper bound given by the minima of $E_\infty^{\rm min}$.}
\label{fig1}
\end{figure}

\begin{figure}[htp!]
\begin{center}
\includegraphics[scale=0.42]{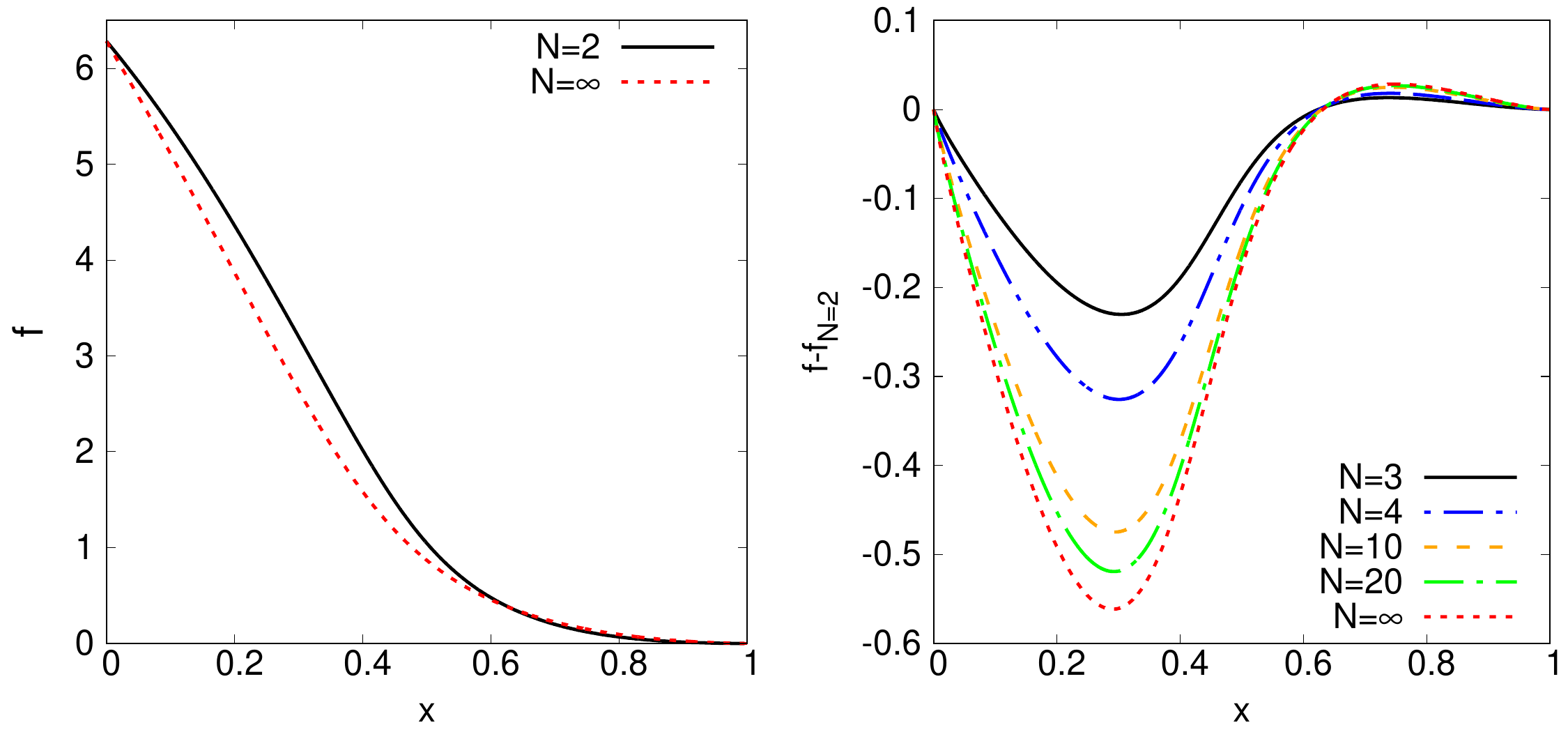}
\end{center}
\caption{The profile functions of the minimizers of \rf{energymin2} for $N=2$ and $N\rightarrow \infty$ are shown on the left. The right picture shows the difference of the profile function $f$ that minimizes \rf{enermin2}, which is a function of $N$, and the profile function for $N=2$.
}
\label{fig2}
\end{figure}

The energies of the $\mid Q \mid =1$  spherically symmetric Skyrmions are shown in Fig. \ref{fig1}, while some of the profile functions are shown in Fig. \ref{fig2}. The lowest energy bound is given by the energy of the $SU(2)$ standard Skyrmion, $E_2^{\rm min} = 1.2314$, while the upper energy bound is given by $E_\infty^{{\rm min}} =  1.5107$. 
In Fig. \ref{fig2}, the plot on the left side shows the profile functions for $N=2,\,\infty$, and on the right side, the difference $f - f_{N=2}$, with $f$ being the profile function associated with the values $N=3,\,4,\,10,\,20,\,\infty$. The shape of the profile functions is slowly deformed as $N$ increases, and the profile functions remain close to each other for different values of $N$, even for the most distinct values. In addition, the modulus of the maximum and minimum values of $f - f_{N=2}$ on the domain $x \in \left[0,\, 1\right]$ increases as $N$ grows.

\section{Conclusion}
\label{sec:conclusion}
\setcounter{equation}{0}

Our work establishes a new and simpler approach for deriving the well-known mathematical result presented in \cite{eells} about the nature of the minimizers of the energy functional \rf{energydef}, now using the idea of self-duality \cite{genbps}. This result states that stable harmonic maps $X$ from the two-sphere $S^2$ to compact Hermitian symmetric spaces $G/{\widehat K} \otimes U(1)$ are holomorphic or anti-holomorphic. Using this result, we propose a generalization of the rational map ansatz in Euclidean space $\IR^3$ for any compact simple Lie group $G$ such that $G/{\widehat K} \otimes U(1)$ is a Hermitian symmetric space, for some subgroup ${\widehat K}$ of $G$.

In the generalized rational map ansatz, the mapping between three-spheres is performed by a radial profile function $f(r)$ and a principal variable $g(z, \bar{z})$ that parametrizes the coset $G/{\widehat K} \otimes U(1)$. Although our ansatz has some limitations, as it is restricted to certain special representations of $G$ where the $S$ matrix \rf{sdef} has the properties given in \rf{omegadef}, we show it works explicitly for the cases of $G=SU(p+q)$ and $G=Sp(2N)$.

Besides providing a class of approximations to topological solutions of the $SU(N)$ Skyrme model \cite{skyrme1, skyrme2, mantonbook}, our ansatz may be used to extend the results for binding energies and rms radii of nuclei obtained by the $SU(2)$ False Vacuum Skyrme model \cite{false} to larger groups. Additionally, this ansatz may be useful in the study of the self-dual sector of a generalized version of the $SU(2)$ quasi-self-dual models proposed in \cite{quasiself} or the $SU(2)$ BPS Skyrme model \cite{laf2017, us1}, also for larger groups. In this last theory, it may lead to the construction of an infinite number of exact topological solutions for each value of the topological charge, as it happens for $G=SU(2)$. The same may be done for the $SU(2)$ quasi-self-dual models proposed in \cite{quasiself}.

\vspace{3cm}

\noindent{\bf Acknowledgements:} The authors are very grateful to Farid Tari, David Brander and F. E. Burstall for very helpful discussions. LAF acknowledges the financial  support of Fapesp (Funda\c c\~ao de Amparo \`a Pesquisa do Estado de S\~ao Paulo) grant  2022/00808-7, and CNPq (Conselho Nacional de Desenvolvimento Cient\'ifico e Tecnol\'ogico) grant 307833/2022-4. LRL acknowledges the financial  support of Fapesp, grant 2022/15107-4. 

\vspace{2cm}

\newpage

\appendix

\section{The homotopic invariance of $Q_{\rm top.}$}
\label{sec:proofq}
\setcounter{equation}{0}

Varying \rf{topchargelambda} one gets
\br
\delta\,Q_{\rm top.}&=&\frac{i}{32\,\pi}\,\int d^2x\, \varepsilon^{ij}\,\left[\trace\(\sbr{\delta\,g\,g^{-1}}{g\,\Lambda\,g^{-1}}\,\sbr{\partial_iX\,X^{-1}}{\partial_jX\,X^{-1}}\)
\right.\nonumber\\
&+& \left. 2\,\trace\(g\,\Lambda\,g^{-1}\,\sbr{\(\partial_i\delta\,X\)\,X^{-1}-\partial_i X\,X^{-1}\,\delta X\,X^{-1}}{\partial_jX\,X^{-1}}\)
\right]
\lab{varytopchargelambda}
\er
Using \rf{dxpdef} and the invariance of the Killing form under the automorphism $\sigma$, we get that the first term in \rf{varytopchargelambda} vanishes
\br
&&\varepsilon^{ij}\,\trace\(\sbr{\delta\,g\,g^{-1}}{g\,\Lambda\,g^{-1}}\,\sbr{\partial_iX\,X^{-1}}{\partial_jX\,X^{-1}}\)
\nonumber\\
&=&
\frac{1}{2}\,\varepsilon^{ij}\,\trace\(\sbr{g^{-1}\,\delta\,g+\sigma\(g^{-1}\,\delta\,g\)}{\Lambda}\sbr{P_i}{P_j}\)=0
\er
since any element of the algebra ${\cal G}$ which is even under $\sigma$ commutes with $\Lambda$. Integrating by parts the first term on the second line of \rf{varytopchargelambda}, one gets
\br
&&\int d^2x\, \varepsilon^{ij}\,\trace\(g\,\Lambda\,g^{-1}\,\sbr{\(\partial_i\delta\,X\)\,X^{-1}}{\partial_jX\,X^{-1}}\)
\nonumber\\
&=&\int d^2x\, \partial_i\left[\varepsilon^{ij}\,\trace\(g\,\Lambda\,g^{-1}\,\sbr{\delta\,X\,X^{-1}}{\partial_jX\,X^{-1}}\)\right]
\nonumber\\
&-&\int d^2x\,\varepsilon^{ij}\,\trace\(\sbr{\partial_ig\,g^{-1}}{g\,\Lambda\,g^{-1}}\,\sbr{\delta\,X\,X^{-1}}{\partial_jX\,X^{-1}}\)
\nonumber\\
&+&\int d^2x\, \varepsilon^{ij}\,\trace\(g\,\Lambda\,g^{-1}\,\sbr{\delta\,X\,X^{-1}\,\partial_i X\,X^{-1}}{\partial_jX\,X^{-1}}\)
\nonumber\\
&+&\int d^2x\, \varepsilon^{ij}\,\trace\(g\,\Lambda\,g^{-1}\,\sbr{\delta\,X\,X^{-1}}{\partial_jX\,X^{-1}\,\partial_i X\,X^{-1}}\)
\lab{longpiece}
\er
Using Gauss' theorem one gets that the first term on the r.h.s. of \rf{longpiece} becomes 
\br
&&\int d^2x\, \partial_i\left[\varepsilon^{ij}\,\trace\(g\,\Lambda\,g^{-1}\,\sbr{\delta\,X\,X^{-1}}{\partial_jX\,X^{-1}}\)\right]
\lab{bcforx}\\
&=&
\int_{S^1_{\infty}} d l_i\, \varepsilon^{ij}\,\trace\(g\,\Lambda\,g^{-1}\,\sbr{\delta\,X\,X^{-1}}{\partial_jX\,X^{-1}}\)\rightarrow 0 \qquad {\rm if} \qquad X=\mbox{\rm constant at}\,S^1_{\infty}
\nonumber
\er
where we have a line integration on a circle $S^1_{\infty}$ at the infinity of the plane $x^1\,x^2$. So, if we assume that the field $X$ goes to a constant at infinity, such a term vanishes. 

Using \rf{longpiece}, \rf{bcforx}, the invariance of the Killing form under $\sigma$, and the antisymmetry of $\varepsilon^{ij}$, one gets that the second line of \rf{varytopchargelambda} becomes
\br
&&\int d^2x\, \varepsilon^{ij}\,\trace\(g\,\Lambda\,g^{-1}\,\sbr{\(\partial_i\delta\,X\)\,X^{-1}-\partial_i X\,X^{-1}\,\delta X\,X^{-1}}{\partial_jX\,X^{-1}}\)
\nonumber\\
&=&-\frac{1}{2}\,\int d^2x\,\varepsilon^{ij}\,\trace\(\sbr{g^{-1}\,\partial_i g+\sigma\(g^{-1}\,\partial_i g\)}{\Lambda}\,\sbr{g^{-1}\,\delta\,X\,X^{-1}\,g}{g^{-1}\,\partial_jX\,X^{-1}\,g}\)
\nonumber\\
&+&\frac{1}{2}\,\int d^2x\, \varepsilon^{ij}\,\trace\(g\,\Lambda\,g^{-1}\,\sbr{\sbr{\delta\,X\,X^{-1}}{\partial_i X\,X^{-1}}}{\partial_jX\,X^{-1}}\)
\nonumber\\
&-&\frac{1}{2}\,\int d^2x\, \varepsilon^{ij}\,\trace\(g\,\Lambda\,g^{-1}\,\sbr{\sbr{\delta\,X\,X^{-1}}{\partial_j X\,X^{-1}}}{\partial_iX\,X^{-1}}\)
\nonumber\\
&+&\frac{1}{2}\int d^2x\, \varepsilon^{ij}\,\trace\(g\,\Lambda\,g^{-1}\,\sbr{\delta\,X\,X^{-1}}{\sbr{\partial_jX\,X^{-1}}{\partial_i X\,X^{-1}}}\)=0
\lab{longpiece2}
\er
The first term on the r.h.s. of \rf{longpiece2} vanishes because $\Lambda$ commutes with any element of the Lie algebra ${\cal G}$ which is even under $\sigma$. The last three terms on the r.h.s. of \rf{longpiece2} add up to zero due to the Jacobi identity of the Lie algebra ${\cal G}$.

Therefore, we have shown that $\delta\,Q_{\rm top.}=0$, i.e. the topological charge \rf{topchargelambda} is invariant under any smooth (homotopic) variation of the fields, as long as the field (principal variable) $X\(g\)$ goes to a constant value at infinity on the plane $x^1\,x^2$.

\section{The calculation of the odd and even matrices and some of their properties} 
\label{sec:appproof}

\subsection{The calculation of $P_i^{(\pm)}$, $\Omega\,P_i^{(\pm)}\,\Omega$, $P_\chi^{(\pm)}$ and $\,P_{\bar{\chi}}^{(\pm)}$}

\setcounter{equation}{0}

From \rf{ws} and \rf{gdef2} we get
\br
P_i &=& +\frac{1}{2}\,\left[g^\dagger\,\(-\frac{\pa_i \vartheta}{\vartheta}\,\(g -\one)\)+i\,\frac{\pa_i S+\pa_i S^\dagger}{\vartheta}+C_i\)\right] \nonumber
\\ 
&& -\frac{1}{2}\,\left[g\,\(-\frac{\pa_i \vartheta}{\vartheta}\,\(g^\dagger -\one)\)-i\,\frac{\pa_i S+\pa_i S^\dagger}{\vartheta}+C_i\)\right] \equiv D_i^- +D_i^+
\er
with $C_i=\frac{1}{\vartheta\,\(1+\vartheta\)}\,\left[\frac{\pa_i\vartheta}{1+\vartheta}\,\(S\,S^\dagger+S^\dagger\,S\)-\(\pa_i S\,S^\dagger+\pa_i S^\dagger\,S+S\,\pa_iS^\dagger+S^\dagger\,\pa_i S\)\right]$ and 
\be D_i^-\equiv \frac{g^\dagger-g}{2}\(\frac{\pa_i\vartheta}{\vartheta}\,\one+C_i\); \qquad\qquad D_i^+\equiv i\,\frac{g^\dagger+g}{2}\,\frac{\pa_i S+\pa_i S^\dagger}{\vartheta}\ee
Note that just the term proportional to $S+S^\dagger$ changes its sign when we take the complex conjugate of $g$ given by \rf{gdef2}. Using also $\omega = \vartheta^2-1$,  $\frac{g^\dagger-g}{2} = -\frac{i}{\vartheta}\(S+S^\dagger\)$, $\frac{g^\dagger+g}{2} = \one - \frac{1}{\vartheta\,(1+\vartheta)}\(S\,S^\dagger +S^\dagger\,S\)$, we obtain  
\br
D_i^-&=& -i\,\frac{\pa_i \vartheta\,\(S+S^\dagger\)}{\vartheta^2}\,\left[1+\frac{\omega}{\(1+\vartheta\)^2}\right] +i \,\frac{S+S^\dagger}{\vartheta^2\,\(1+\vartheta\)}\,\(\pa_i S\,S^\dagger+\pa_i S^\dagger\,S+S\,\pa_iS^\dagger+S^\dagger\,\pa_i S\)\nonumber\\
&=&\frac{i}{\vartheta^2\,\(1+\vartheta\)}\left[-\pa_i \omega\,\(S+S^\dagger\) +S\,\pa_i S^\dagger\,S+S\,S^\dagger\,\pa_i S + S^\dagger\,\pa_i S\,S^\dagger+S^\dagger\,S\,\pa_iS^\dagger
\right] \lab{d-}\\
D_i^+ &=& \frac{i}{\vartheta^2\,\(1+\vartheta\)}\left[\vartheta\,\(1+\vartheta\)\,\(\pa_i S+\pa_i S^\dagger\)-S\,S^\dagger\,\pa_i S-S^\dagger\,S\,\pa_i S^\dagger\right] \lab{d+}
\er
Therefore,
\br
P_i &=& P_i^{(+)}+P_i^{(-)};\qquad {\rm with} \qquad  P_i^{(+)}= \frac{i}{\vartheta}\,\left[\pa_iS+\frac{S\,\pa_i S^{\dagger}\, S-\pa_i \omega\,S }{\vartheta\,\(1+\vartheta\)} \right]\nonumber\qquad \\ P_i^{(-)} &=& \frac{i}{\vartheta}\,\left[\pa_iS^{\dagger} +\frac{S^{\dagger}\,\pa_i S\, S^{\dagger}-\pa_i \omega \,S^\dagger}{\vartheta\,\(1+\vartheta\)} \right] \lab{psapp}
\er
which satisfies $(P_i^{(+)})^\dagger = - P_i^{(-)}$. 

 Let us introduce the hermitian and invertible operators 
\br
A &\equiv &\one + \frac{1}{1+\vartheta}\,S\,S^\dagger \qquad\qquad \Rightarrow \qquad\qquad A^{-1} =\one - \frac{1}{\vartheta\(1+\vartheta\)}\,S\,S^\dagger \nonumber\\
B &\equiv &\one + \frac{1}{1+\vartheta}\,S^\dagger\,S \qquad\qquad \Rightarrow \qquad\qquad B^{-1} =\one - \frac{1}{\vartheta\(1+\vartheta\)}\,S^\dagger \,S \lab{ab}
\er
Its so follows from $1+\frac{\omega}{1+\vartheta}=\vartheta$, \rf{psapp} and \rf{ab} that
\br
A\,P_i^{(+)} &=& \frac{i}{\vartheta}\,\left[\pa_iS+\frac{1}{\vartheta\(1+\vartheta\)}\, \(\(S\,\pa_i S^{\dagger}\, S-\pa_i \omega\,S\)\,\(1+\frac{\omega}{1+\vartheta}\) +\vartheta\,S\,S^{\dagger}\, \pa_i S\) \right] \nonumber\\ 
&=& \frac{i}{\vartheta}\,\left[\pa_iS+\frac{1}{1+\vartheta}\, \(S\,\pa_i S^{\dagger}\, S-\pa_i \omega\,S +S\,S^{\dagger}\, \pa_i S\) \right]
\er
Thus,
\br
A\,P_i^{(+)}\,B &=& \frac{i}{\vartheta}\,\left[\pa_iS +\frac{1}{1+\vartheta}\, \(S\,\pa_i S^{\dagger}\, S-\pa_i \omega\,S +S\,S^{\dagger}\, \pa_i S\) \right] \nonumber \\
&& +\frac{i}{\vartheta\,\(1+\vartheta\)}\,\left[\pa_iS\,S\,S^\dagger +\frac{1}{1+\vartheta}\, \(\omega\(S\,\pa_i S^{\dagger}\, S-\pa_i \omega\,S\) +S\,S^{\dagger}\, \pa_i S\,S^\dagger\,S\) \right] \nonumber\\
&=&\frac{i}{\vartheta}\,\left[\pa_iS +\frac{1}{1+\vartheta}\, \(\pa_i\( S\, S^{\dagger}\, S\)-\pa_i \omega\,S \) \right] \nonumber \\
&& +\frac{i}{\vartheta\,\(1+\vartheta\)}\,\left[\frac{1}{1+\vartheta}\, \(\omega\(S\,\pa_i S^{\dagger}\, S-\pa_i \omega\,S\) +S\,S^{\dagger}\, \pa_i S\,S^\dagger\,S\) \right] \lab{apb}
\er
However, from \rf{omegadef} we can obtain some useful relations, such as 
\be \pa_i S\,S=\pa_i S^\dagger\,S^\dagger=0;\qquad\qquad S\,S^\dagger\,\pa_i S\,S^\dagger\,S =\omega\,\(\pa_i \omega \,S-S\,\pa_iS^\dagger\, S\) \lab{usefulrel1}\ee
which reduces \rf{apb} to
\be A\,P_i^{(+)}\,B = i\,\pa_i S \;\qquad\qquad \stackrel{\dagger}{\Rightarrow}\qquad\qquad B\,P_i^{(-)}\,A = i\,\pa_i S^\dagger \lab{extraction0}\ee
where we use $(P_i^{(+)})^\dagger = - P_i^{(-)}$. In addition, from \rf{omegadef}, \rf{psapp} and \rf{ab}  we also have that
\be S\,P_i^{(+)} = P_i^{(+)}\,S= 0\;;\qquad\qquad \qquad\qquad S^\dagger\,P_i^{(-)} = P_i^{(-)}\,S^\dagger= 0 \lab{inv}\ee
Therefore,  using \rf{ab}, \rf{extraction0} and \rf{inv} the invertible and hermitian operator
\be \Omega \equiv \one + \frac{S\,S^\dagger+ S^\dagger\,S}{1+\vartheta} \qquad \Rightarrow \qquad \Omega^{-1} =\one - \frac{S\,S^\dagger+S^\dagger\,S}{\vartheta\,\(1+\vartheta\)} \ee
satisfies
\be \Omega\,P_i^{(+)}\,\Omega = i\,\pa_i S \;\qquad\qquad \stackrel{\dagger}{\Rightarrow}\qquad\qquad \Omega\,P_i^{(-)}\,\Omega = i\,\pa_i S^\dagger \lab{extraction2}\ee
Therefore, choosing an ansatz such that $\pa_{\bar{\chi}} S=0$ is equivalent to imposing  $P_{\bar{\chi}}^{(+)}=0$, where $\chi$ can be either $z$ or $\bar{z}$. In particular, the fact that $\partial_{\bar{\chi}} S=0$ implies  $P_{\bar{\chi}}^{(+)}=0$  can also be easily demonstrated by directly computing $P_{\bar{\chi}}$ using \rf{psapp} and $S\,\pa_{\bar{\chi}} S^{\dagger}\, S=\pa_{\bar{\chi}} \omega\,S$. Consequently, $P_{{\chi}}^{(-)}=0$ and the non-vanishing components of  \rf{psapp} become
\br
P_\chi^{(+)} &=& i\,\frac{\(1+\vartheta\)^2}{\vartheta}\,\pa_\chi \(\frac{S}{\(1+\vartheta\)^2}\);\qquad\quad P_{\bar{\chi}}^{(-)}= i\,\frac{\(1+\vartheta\)^2}{\vartheta}\,\pa_{\bar{\chi}} \(\frac{S^\dagger}{\(1+\vartheta\)^2}\) \lab{pirapp}\er

\subsection{The calculation of $K_i$, $K_\chi$ and $K_{\bar{\chi}}$}

From \rf{ws} and \rf{gdef2} we get
\br
K_i &=& +\frac{1}{2}\,\left[g^\dagger\,\(-\frac{\pa_i \vartheta}{\vartheta}\,\(g -\one)\)+i\,\frac{\pa_i S+\pa_i S^\dagger}{\vartheta}+C_i\)\right] \nonumber
\\ 
&& +\frac{1}{2}\,\left[g\,\(-\frac{\pa_i \vartheta}{\vartheta}\,\(g^\dagger -\one)\)-i\,\frac{\pa_i S+\pa_i S^\dagger}{\vartheta}+C_i\)\right] 
\er
with $C_i=\frac{1}{\vartheta\,\(1+\vartheta\)}\,\left[\frac{\pa_i\vartheta}{1+\vartheta}\,\(S\,S^\dagger+S^\dagger\,S\)-\(\pa_i S\,S^\dagger+\pa_i S^\dagger\,S+S\,\pa_iS^\dagger+S^\dagger\,\pa_i S\)\right]$ and 
\br
K_i &=& \frac{1}{2}\,\left[\frac{\pa_i \vartheta}{\vartheta}\,\(-2\,\one +g+g^\dagger \) +i\,\(g^\dagger - g\)\,\(\frac{\pa_i S+\pa_i S^\dagger}{\vartheta}\) + \(g^\dagger + g\)\, C_i \right]  \,\nonumber\\
&= & K^1_i + K^2_i + K^3_i  \lab{123} \er
with 
\br 
K_i^1 &\equiv & \frac{1}{2}\,\frac{\pa_i \vartheta}{\vartheta}\,\(-2\,\one +g+g^\dagger \)  = - \frac{\pa_i \vartheta}{\vartheta^2}\,\frac{1}{\vartheta+1} \,\left(S\,S^\dagger+S^\dagger\,S\right)  \nonumber
\\
K_i^2 &\equiv & \frac{i}{2}\,\(g^\dagger - g\)\,\(\frac{\pa_i S+\pa_i S^\dagger}{\vartheta}\) = \frac{1}{\vartheta}\, \(S+S^\dagger\) \,\(\frac{\pa_i S+\pa_i S^\dagger}{\vartheta}\) = \frac{1}{\vartheta^2}\,\(S^\dagger\,\pa_i S+S\,\pa_i S^\dagger\) \nonumber
\\
K_i^3 &\equiv & \frac{g^\dagger + g }{2}\, C_i =   \left[ \one -\frac{1}{\vartheta\(\vartheta+1\)} \,\left(S\,S^\dagger+S^\dagger\,S\right) \right] \, C_i = C_i -\frac{1}{\vartheta^2\,\(1+\vartheta\)^2}\,\cdot \nonumber \\
&  & \cdot \left[\frac{\pa_i\vartheta\,\omega}{1+\vartheta}\,\(S\,S^\dagger+S^\dagger\,S\)-\(S\,S^\dagger\,\pa_i S\,S^\dagger+S^\dagger\,S\,\pa_i S^\dagger\,S+\omega\,\(S\,\pa_iS^\dagger+S^\dagger\,\pa_i S\)\)\right]\nonumber
\er
Using  $\(1-\frac{\omega}{\vartheta\,(\vartheta+1)}\)=\frac{\vartheta\,(\vartheta+1)-(\vartheta^2-1)}{\vartheta\,(\vartheta+1)}=\frac{1}{\vartheta}$ we obtain
\br 
K_i^3 &\equiv & \frac{1}{\vartheta^2\,\(1+\vartheta\)}\,\left[\frac{\pa_i\vartheta}{1+\vartheta}\,\(S\,S^\dagger+S^\dagger\,S\)-S\,\pa_iS^\dagger - S^\dagger\,\pa_i S\right] + k^3_i\nonumber \er 
with
\br k^3_i & \equiv & -\frac{1}{\vartheta\,\(1+\vartheta\)}\,\left[\pa_i S\,S^\dagger + \pa_i S^\dagger\,S -\frac{1}{\vartheta\(\vartheta+1\)} \,\(S\,S^\dagger\,\pa_i S\,S^\dagger+S^\dagger\,S\,\pa_i S^\dagger\,S\)\right]\nonumber
\er
Therefore, 
\br
K_i^1+K_i^3 & = & \frac{1}{\vartheta^2\,\(1+\vartheta\)}\,\left[-\frac{\vartheta\,\pa_i\vartheta}{1+\vartheta}\,\(S\,S^\dagger+S^\dagger\,S\)-S\,\pa_iS^\dagger - S^\dagger\,\pa_i S\right] + k^3_i\nonumber 
\er
Using also $\vartheta\,\pa_i \vartheta = \frac{1}{2}\,\pa_i \omega$, the quantity \rf{123} becomes
\br
K_i & = & \frac{1}{\vartheta\,\(1+\vartheta\)}\,\left[-\frac{1}{2}\,\frac{\pa_i\omega}{\vartheta\(1+\vartheta\)}\,\(S\,S^\dagger+S^\dagger\,S\)+S\,\pa_iS^\dagger + S^\dagger\,\pa_i S\right] + k^3_i\lab{kismall}
\er

Note that $S\,S^\dagger\,\pa_\chi S\,S^\dagger = S\,\pa_\chi\(S^\dagger\, S\,S^\dagger \) = \pa_\chi\omega \, S\,S^\dagger$ and $S^\dagger\, S\, \pa_{\bar{\chi}} S^\dagger\,S = \pa_{\bar{\chi}}\omega \,S^\dagger \, S$, then $k^3_\chi = -\frac{1}{\vartheta\,\(1+\vartheta\)}\,\left[\pa_\chi S\,S^\dagger  -\frac{1}{\vartheta\(\vartheta+1\)} \,\pa_\chi\omega \, S\,S^\dagger\right]$ and $k^3_{\bar{\chi}} = -\frac{1}{\vartheta\,\(1+\vartheta\)}\,\left[\pa_{\bar{\chi}} S^\dagger \, S  -\frac{1}{\vartheta\(\vartheta+1\)} \,\pa_{\bar{\chi}}\omega \,S^\dagger \, S\right]$. Thus, the components of \rf{kismall} become
\br
K_\chi & = & + \frac{1}{\vartheta\,\(1+\vartheta\)}\,\left[\frac{1}{2}\,\frac{\pa_\chi \omega}{\vartheta\(1+\vartheta\)}\,\left[S ,\, S^\dagger\right] - \pa_\chi \left[S ,\, S^\dagger\right]\right]   \\
K_{\bar{\chi}} & = & - \frac{1}{\vartheta\,\(1+\vartheta\)}\,\left[\frac{1}{2}\,\frac{\pa_{\bar{\chi}} \omega}{\vartheta\(1+\vartheta\)}\,\left[S ,\, S^\dagger\right] - \pa_{\bar{\chi}} \left[S ,\, S^\dagger\right]\right]  
\er

\end{document}